%% file: acl2023.tex
\pdfoutput=1

\documentclass[11pt]{article}

\usepackage{ACL2023}
\usepackage{times}
\usepackage{latexsym}
\usepackage{amsmath}
\usepackage{graphicx}
\usepackage{booktabs}
\usepackage{amsfonts}
\usepackage{multirow}
\usepackage{enumerate}
\usepackage{tcolorbox}
\tcbuselibrary{breakable}
\usepackage{amsmath}
\usepackage{wrapfig}
\usepackage{algorithm}
\usepackage{algorithmic}

\usepackage{inconsolata}
\usepackage{subfigure}

\usepackage[T1]{fontenc}

\usepackage[utf8]{inputenc}

\usepackage{microtype}

\title{Unveiling the Truth and Facilitating Change: Towards Agent-based
Large-scale Social Movement Simulation}

 \author{
    Xinyi Mou$^1$,
    Zhongyu Wei$^{1,2}$\Thanks{ Corresponding author.},
    Xuanjing Huang$^{3,4}$\\
    \textsuperscript{\rm 1}School of Data Science, Fudan University, China\\
    \textsuperscript{\rm 2}Research Institute of Intelligent and Complex Systems, Fudan University, China\\
    \textsuperscript{\rm 3}School of Computer Science, Fudan University, China\\
    \textsuperscript{\rm 4}Shanghai Collaborative Innovation Center of Intelligent Visual Computing, China\\
    \texttt{\{xymou20,zywei,xjhuang\}@fudan.edu.cn}
    \ \ \ 
    }

\begin{document}
\maketitle
\begin{abstract}
Social media has emerged as a cornerstone of social movements, wielding significant influence in driving societal change. Simulating the response of the public and forecasting the potential impact has become increasingly important. However, existing methods for simulating such phenomena encounter challenges concerning their efficacy and efficiency in capturing the behaviors of social movement participants. In this paper, we introduce a hybrid framework \textbf{HiSim} for social media user simulation, wherein users are categorized into two types. Core users are driven by Large Language Models, while numerous ordinary users are modeled by deductive agent-based models. We further construct a Twitter-like environment to replicate their response dynamics following trigger events. Subsequently, we develop a multi-faceted benchmark \textbf{SoMoSiMu-Bench} for evaluation and conduct comprehensive experiments across real-world datasets. Experimental results demonstrate the effectiveness and flexibility of our method \footnote{Code and data are available at \url{https://github.com/xymou/social_simulation}.}.

\end{abstract}

\section{Introduction}
\begin{figure}[!t]
    \centering
    \includegraphics[width=0.99 \linewidth]{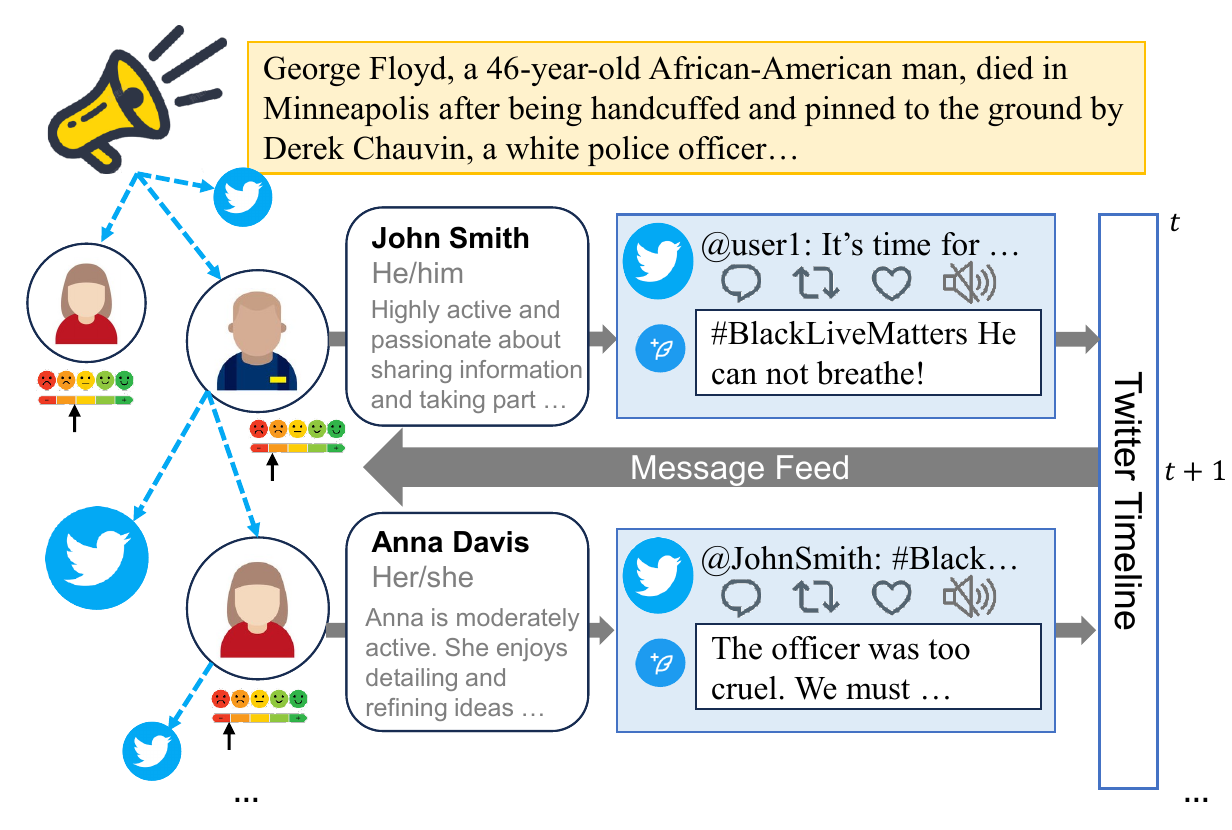}
    \caption{An illustration of user interactions and attitude changes after a trigger event happens. Users can take actions such as posting and retweeting according to their traits, and their generated content will be stored in the Twitter timeline and fed to their connected users. Users can change attitudes once perceive others' opinions.}
    \label{img:intro}
\end{figure}

In the past decades, social media has witnessed many social movements, such as the Arab Spring~\cite{rane2012social} and \#Metoo~\cite{brunker2020role}. Twitter stands out as a prominent forum giving powerful voices to groups demanding change. As illustrated in Figure~\ref{img:intro}, the dissemination of breaking news on Twitter prompts the proliferation of opinions, influencing collective sentiment and shaping societal agendas, often resulting in real-world actions~\cite{roy2023tale}. Although the majority of social movements are reported peaceful, the sheer scale of participation can sometimes escalate into violence and destruction, posing potential ramifications. Therefore, proactive measures to anticipate the impact of such events become imperative. 

Previous research on analyzing online social movements has primarily concentrated on retrospective analysis of content and users~\cite{blmdata,roy2023tale}, rather than utilizing simulation for prediction. Agent-based models (ABMs) have been extensively employed for simulation in social science~\cite{schelling2006micromotives, jackson2017agent}, wherein each agent symbolizes an individual, and interactions among agents give rise to distinct social phenomena. Typically, ABMs are micro-level mathematical models that define how individuals affect each other, creating collective social patterns through simulating interaction at scale~\cite{tornberg2023simulating}.

Recently, Large Language Models (LLMs) have demonstrated impressive ability in human-level intelligence~\cite{wang2023survey, xi2023rise}. LLM-based user simulations have been successfully experimented in domains such as recommendation~\cite{wang2023recagent, zhang2023generative} and collaborative work~\cite{chen2023agentverse, qian2023experiential}. However, the exploration of conducting large-scale online social movement simulations using LLMs remains limited and presents the following challenges: (1) How to accurately simulate users of social media and replicate their behaviors within the community? (2) How to efficiently simulate a large number of users, given the impracticality of employing thousands of LLMs? (3) How to comprehensively evaluate the effectiveness of the simulation?

To handle these challenges, this paper introduces \textbf{HiSim}: a novel \textbf{\underline{h}ybrid framework for soc\underline{i}al media user \underline{sim}ulation}. Considering the Pareto distribution~\footnote{\url{https://en.wikipedia.org/wiki/Pareto_distribution}} inherent in social media user engagement, we categorize users into two types: core users, comprising active and influential figures such as opinion leaders, and ordinary users. Core users are characterized and driven by LLMs, enabling emulation of their complex behaviors, while massive ordinary users are governed by ABMs, providing a practical way for user simulation in large scale.

Based on the hybrid mechanism composed of two types of users, we establish an \textbf{online social media environment} tailored for online social movement simulation and evaluation. In this environment, messages are organized in Twitter-like timelines and offline news can be disseminated. User interactions and resultant collective attitudes are observed through attitude scores. To systematically evaluate the simulation, we propose a novel benchmark \textbf{SoMoSiMu-Bench}, including three real-world collected datasets (\textit{Metoo}, \textit{RoeOverturned} and \textit{BlackLivesMatter}) and an evaluation strategy at both the micro and macro levels, focusing on individual user alignment and systemic outcomes respectively. Evaluation results on SoMoSiMu-Bench demonstrate the effectiveness of our simulation framework.

Our contributions can be summarized as follows:
\begin{enumerate}[-]
\itemsep-0.2em
    \item We introduce a hybrid simulation framework where two types of users are separately modeled, to tackle the cost and efficiency challenges associated with simulating massive participants. 
    \item We develop a simulator tailored for online social movements, featuring a Twitter-like environment and modeling of user opinion dynamics.
    \item We provide the first benchmark SoMoSiMu-Bench for social movement simulation evaluation, including a data collection consisting of three real-world movements and corresponding evaluation methods. Experiment results and analysis demonstrate the effectiveness of our method.
\end{enumerate}

\begin{figure*}[!ht]
    \centering
    \includegraphics[width= \linewidth]{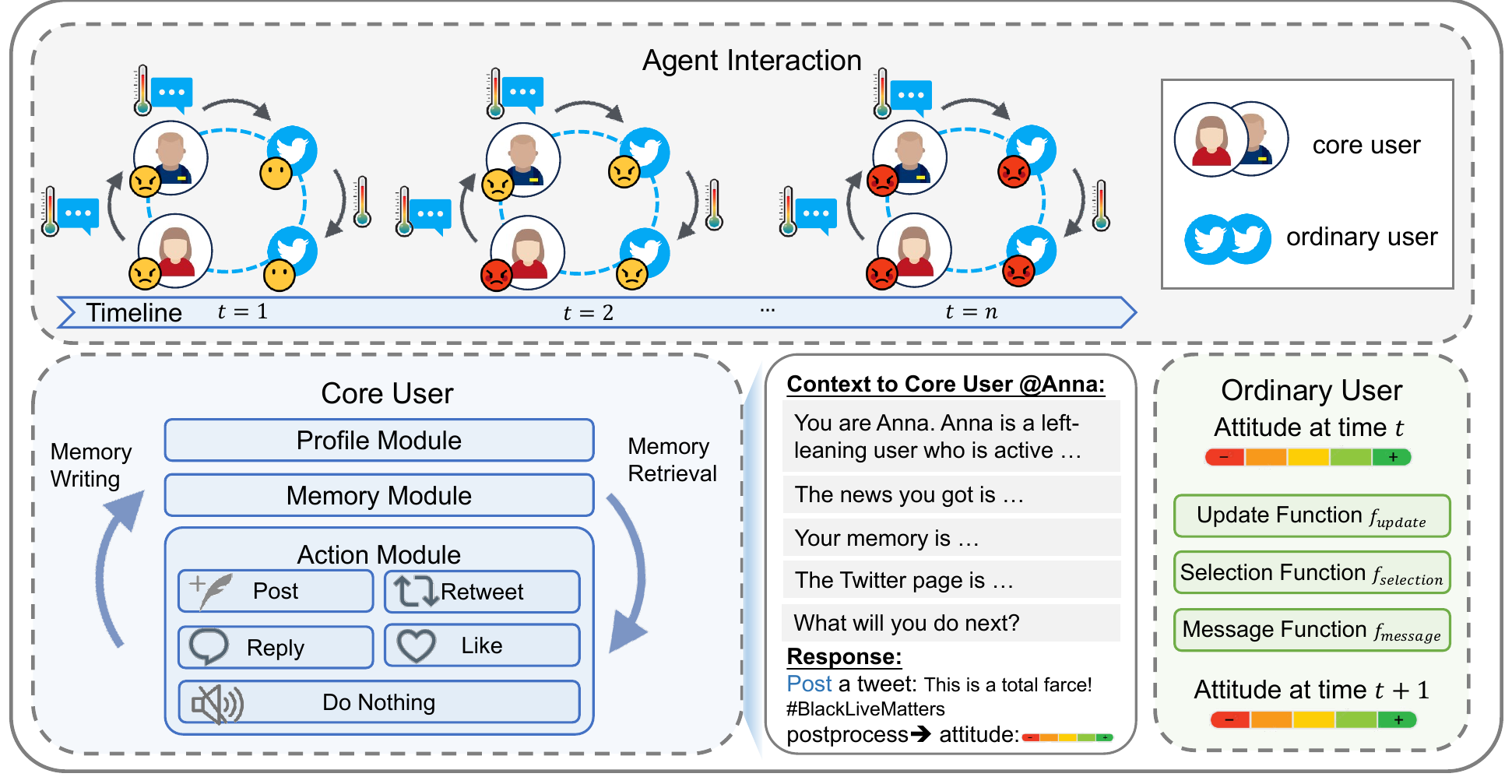}
    \caption{The proposed framework architecture. The bottom part illustrates the architecture of core users and the mechanism for ordinary users. The top part presents the simulation process. At each round, core user agents take action by generating textual responses based on contextual information, and their attitudes are conveyed to ordinary users after postprocessing, while ordinary users communicate using attitude scores directly.}
    \label{img:fm}
\end{figure*}

\section{Formalization of Public Opinion Dynamics Simulation}
Modeling the change in people's attitudes, beliefs and opinions is crucial since opinion change can result in societal phenomena such as bipolarization and extremization. In this section, we present the preliminaries of agent-based models and the formulation of the targeted task.

\subsection{Preliminaries}
\subsubsection{Agent-based Models in Opinion Dynamics}\label{sec:abm}
Agent-based models (ABMs) are micro-level mathematical models defining how an individual agent adjusts the attitudes toward specific topics based on the opinions of others~\cite{lorenz2021individual,chuang2023computational}. By simulating interactions among individual behaviors, ABMs enable the identification of emergent, group-level mechanisms that could not be predicted using the characteristics of individuals within a social system alone~\cite{tornberg2023simulating}.

Typically, in ABMs for opinion dynamics, each agent keeps a continuous attitude score representing its opinion, where the sign of the score represents attitude direction, i.e., positive or negative, and the magnitude of the score describes the attitude intensity. ABMs define how this score is changed under the influence of others. Although the exact formulations vary in different ABMs, most models can be decomposed into components that are present across ABMs and be expressed in a unified formulation~\cite{chuang2023computational}, where three key functions can specify it, i.e., the \textit{attitude update function} $f_{\text{update}}$, the \textit{selection function} $f_{\text{selection}}$ and the \textit{message function} $f_{\text{message}}$. 

\paragraph{Update Function}
The update function generally defines the change of attitudes. Formally, the attitude update is:
\begin{align}
    \Delta a_{i, t}=a_{i, t+1}-a_{i, t}=f_{\text {update }}\left(a_{i, t}, M_{i, t}\right),
\end{align}
where $\Delta a_{i,t}$ is the attitude change of agent $i$ from time step $t$ to $t+1$, $a_{i,t}$ and $a_{i,t+1}$ are the attitude of agent $i$ before and after interaction, and $M_{i, t}=\left\{m_{j, t} \mid j \in J_{i, t}\right\}$ are the messages the agent $i$ receive from $J_{i,t}$, i.e., those who interact with the agent $i$ at time step $t$.

\paragraph{Selection Function}
The selection function determines the set of agents $J_{i,t}$ that will have a social influence on agent $i$. It can be driven by internal factors like internal intendancy to interact with those more similar to them, or by external factors such as recommendation algorithms of the platforms.

\paragraph{Message Function}
The message function determines the message $m_{j,t}$ that agent $j$ shares based on its attitude $a_{j,t}$. Formally, it is also a continuous score, a function of the attitude $a_{j,t}$:
\begin{align}~\label{eq3}
    m_{j, t}=f_{\text {message }}\left(a_{j, t}\right),
\end{align}
Most ABMs assume that agents convey their internal attitude with other agents without bias, i.e., $ m_{j, t}=f_{\text {message }}\left(a_{j, t}\right)=a_{j,t}$.

\subsubsection{LLM-empowered Agents}
Recently, a growing research area employs LLMs to construct autonomous agents, where the key idea is to equip LLMs with crucial human capabilities such as memory and planning. In terms of agent architecture, existing research can be synthesized into a unified framework consisting of the profile module, the memory module, the planning module, and the action module~\cite{wang2023survey}. They are designed to indicate the profiles of the agent roles, help the agents accumulate experiences and self-evolve, deconstruct complex tasks, and translate the agent's decisions into specific outcomes. Benefiting from the strong generative ability of LLMs, LLM-empowered agents can model more complex behaviors of users instead of simply representing opinions with a single score.

\subsection{Task Formulation}\label{sec:task}
 In this paper, we aim to predict how a group of users' opinions on a social movement event change through agent-based simulation, and compare the simulation results with real-world scenarios. We consider a group of users $\mathcal{U}=\{1, \ldots, U\}$, each of whom participates in an online social movement and has an attitude on the specific topic. The attitudes evolve through social interaction. Let $a_{i, t} \in \mathbb{A} = [-1, 1]$ be the attitude that user $i$ holds at time step $t \in\{1,2, \ldots\}=\mathbb{N}$, where the sign of $a_{i,t}$ entails the direction of the attitude and the absolute value of $a_{i,t}$ represents the magnitude of the attitude. For each user, we instantiate the corresponding agent with the user's initial attitude and profile, and construct the social networks based on the authentic following relationships on Twitter. Then, we aim to (1) simulate the behaviors of users at the individual level given a certain context in the pattern of single-round simulation; and (2) simulate continuously to observe how collective opinions shift over time resulting from user interactions.

\section{Hybrid Framework for Social Media Simulation}

User engagement in social networks often exhibits a Pareto distribution, where the bulk of content originates from a small fraction of individuals. Thus, those more active and influential such as opinion leaders should be modeled finely, while the silent majority can be controlled by simpler models. The overall framework is illustrated in Figure~\ref{img:fm}, where social media users are divided into core users and ordinary users. The two types of users are driven by different models, to address the cost and efficiency issues of using thousands of LLMs.

\subsection{Simulation of Core Users}
We build an agent architecture by empowering LLMs with the necessary capabilities for core user simulation. An overview of the agent's architecture is illustrated in the left lower part of Figure~\ref{img:fm}. The agent is equipped with a profile module, a memory module, and an action module, to complete the complex operations on Twitter.

\subsubsection{Profile Module}\label{sec:profile}
We extract and summarize the following information from real user data and prompt the corresponding agents when simulating:
\paragraph{Demographics}
The basic profile is demographics, such as name, gender,  political leaning and account type~\cite{brunker2020role}. This information is highly related to the user's potential stance on social events. We induce the demographics from users' biographies and previous tweets. The implementation details can be found in Appendix~\ref{app:profile}.

\paragraph{Social Traits}
As participants in a social platform, agents' social traits such as activity and influence also capture important characteristics. Activity quantifies the frequency of a user's interaction, while influence reflects the quality and popularity of generated content. Since users exhibit long-tail distribution among these social traits, we segment them into three uneven tiers~\cite{zhang2023generative}.

\paragraph{Communication Roles}
To more accurately describe users in participation of social movements, we integrate Edelman's topology of influence (TOI) ~\cite{bentwooddistributed, tinati2012identifying} to identify the communication roles of online users: (1) \textbf{Idea Starter}: individuals who start the conversation and post original content. (2) \textbf{Amplifier}: users who collect multiple thoughts and share ideas and opinions. (3) \textbf{Curator}: they use a broader context to define ideas. They tend to take the ideas of others and either validate, question, challenge or dismiss them. (4) \textbf{Commentator}: users who take part in something to which they strongly feel about. They retweet actively. (5) \textbf{Viewer}: the inactive majority, who prefer to consume information rather than create or share information online.

\subsubsection{Memory Module}
We consider two types of memory to fully characterize the social media user and reconstruct the human-like memory mechanisms. (1) \textbf{Personal Experience}: The personal experience is authentic records of the users, which can be extracted from the users' historical tweets before the event happens. By retrieving the relevant experience and opinions of the user, it would be easier to infer how this user would behave in similar situations. (2) \textbf{Event Memory}: The event memory represents the observation of the agent itself and other visible agents. It captures specific and concentrated insights after the event happens, i.e., after the simulation starts. We integrate a memory module to manipulate the memories of agents, mainly including three operations:
\paragraph{Memory Writing}
The raw observations including behaviors performed by the agents themselves and tweets visible to the agents are input into the memory module after each round's interaction, in both forms of natural languages and vectors.

\paragraph{Memory Retrieval} Agents can extract information from the memories considering different factors. The retrieval function gets observations based on recency, relevance, importance and immediacy ~\cite{park2023generative, chen2023agentverse}, where recency assigns a higher score to memory objects that were recently accessed, relevance assigns a higher score to memory objects that are related to the current situation, importance assigns a higher score to memory objects that the agent believes to be important, and immediacy assigns a higher score to memory that needs quick attention or immediate response. The top-ranked memories are subsequently integrated as part of the prompt.

\paragraph{Memory Reflection}
We incorporate the reflection operation to urge the agents to generate high-level thoughts. We follow ~\citeauthor{park2023generative} to implement reflection periodically, with steps including: (1) generating the most salient questions that can be asked given the agent's recent experiences; (2) prompting agents to extract high-level insights from retrieved relevant memories. This type of memory will be included alongside other observations when retrieval occurs.

\subsubsection{Action Module}
We design an action module tailored for social media ecology, where actions are highly related to information and attitude propagation, including: (1) \textbf{Post}: post original content; (2) \textbf{Retweet}: retweet an existing tweet in the agent's page, either forward directly or post additional statements; (3) \textbf{Reply}: reply to authors of existing tweets or replies; (4) \textbf{Like}: like an existing tweet; (5) \textbf{Do Nothing}: do nothing and keep silent. The optional actions are presented to the agents via prompting. The agents' responses are then parsed into concrete effects on the environment, such as adding a new tweet or increasing the retweets of an existing tweet.

\subsection{Simulation of Ordinary Users}
\paragraph{Initial Attitudes}
To restore the real situation and lay the foundation for reliable simulation, we initialize the attitudes based on the corresponding user’s tweets at that time period, instead of setting the initial opinions uniformly distributed. This can be implemented by annotating the direction and density of their generated content on Twitter.

\paragraph{Attitude Change Mechanism}
We employ ABMs in opinion dynamics in Sec.~\ref{sec:abm} to model the attitude change of ordinary users. Formally, at time step $t$, agent $i$ interacts with a set of agents $J_{i,t}$ based on the selection function $f_{\text{selection}}$. The selected agents then share their messages based on the message function $f_{\text{message}}$, which is a function of their attitudes. After receiving the messages, agent $i$ updates its attitude from $a_{i,t}$ to $a_{i,t+1}$ based on the attitude update function $f_{\text{update}}$.

\subsection{Interaction between Agents}
In the hybrid system, the interaction between different agents is shown at the top of Figure~\ref{img:fm}. 
\paragraph{Interaction between Homogeneous Agents}
Core users convey their thoughts to others by generating specific content, in the form of natural languages.
For example, as shown in Figure~\ref{img:fm}, user Anna generates a post at time step $t$, and this content will be part of the "Twitter page" in prompt at time step $t+1$ for other users who follow Anna. For ordinary users, information is transmitted according to the message function defined in ABMs.

\paragraph{Interaction between Heterogeneous Agents}
Since ABMs only accept numeric inputs and outputs, we need to transform the content generated by core users into attitude scores for ABMs. External LLMs are employed to annotate the stance, i.e., attitude direction of the content, and sentiment analysis tool is applied to calculate the attitude intensity. After this postprocessing, the scores can be processed by the message function in ABMs. Considering that the impact of ordinary users on core users is subtle, we currently do not address the influence from ordinary users to core users.

\subsection{Simulation Environment}
To simulate and evaluate users' reactions during real events, we build a Twitter-like simulation platform that the agents are situated within and discuss the execution of the simulation.

\paragraph{Message Feed Mechanism}
The environment operates based on the concept of timeline~\cite{tinati2012identifying}. In this environment, each user has a timeline of tweets created by themselves and other users they follow. Also, a public timeline is kept to store tweets sent by all users. At each round, the most recent tweets are provided for prompting.

\paragraph{Offline News Feed}
Some offline events often act as catalysts for social movements, such as the George Floyd incident triggering the widespread \textit{\#BlackLivesMatter} movement. Thus, we provide real-world events described in natural languages as background information to the core user agents.

\subsection{Simulation Process}
Our simulator operates in different ways for different purposes in Sec.~\ref{sec:task}. To validate the replication of user behaviors, the simulator can run in a single round, where the provided context is authentic. To estimate future public opinion, our simulator can also operate in a round-by-round manner, where the subsequent context contains simulated content. During each round, i.e., time step, agents for core users autonomously give a thought before taking actions and then decide what actions they would like to take. Overall, LLM agents for core users perform actions based on the following information: (1) profile or description of the agent; (2) memory of the agent; (3) triggering offline news; (4) the Twitter page showing tweets visible to the agent; (5) notifications containing replies to the agent. A full prompt example can be found in Appendix~\ref{app:prompt4user}. Agents for ordinary users update their attitudes based on the pre-defined formulas in ABMs and perceived messages from other agents. The process is shown in the algorithms in Appendix~\ref{app:alg}.

\section{SoMoSiMu-Bench: A Benchmark for Social Movement Simulation}
In this section, we present the \textbf{SoMoSiMu-Bench}, a benchmark for simulation evaluation. We construct a data collection, composed of three social movements on Twitter. Then, evaluation strategies at the micro and macro levels are designed.

\input{tabs/data}

\subsection{Datasets}~\label{sec:4.1}
We first present the construction of our dataset.

\paragraph{Data Collection}
To broadly evaluate the simulation performance of the proposed method, we construct three Twitter datasets by collecting tweets related to specific social movements, i.e., \textit{Metoo}~\cite{metoodata}, \textit{RoeOverturned (Roe)}~\cite{chang2023roeoverturned} and \textit{BlackLivesMatter (BLM)}~\cite{blmdata}. For each movement, we collect tweets spanning two specific events or phases, as outlined in Table~\ref{tab:dataset}.

\paragraph{User Selection}
Due to the absence of an authoritative definition for core users, we identify core users by ranking all participants based on the activity and influence metrics in practice. 
From the gathered tweets, we select 300 core users by first identifying the top 100 most influential individuals based on the number of received retweets. We then extend this selection by selecting an additional 200 active users from their social networks, based on their overall tweet frequency. Next, we randomly sample ordinary users from those who tweeted during the event period. Subsequently, we collect their social networks and tweets during the event period and annotate the attitude scores using GPT-3.5~\cite{gpt3.5} and Textblob~\footnote{\url{https://github.com/sloria/TextBlob}}. 

To reduce the \emph{annotation cost} for validation, rather than the \emph{simulation cost}, we retain 700 ordinary users. As a result, 1,000 users are acquired for the simulation of each event. The statistics of the datasets are presented in Table~\ref{tab:dataset}. More details can be found in Appendix~\ref{app:data}.

\subsection{Micro Alignment Evaluation}
To evaluate the effectiveness of the simulation at the individual level, we simulate in single rounds by providing authentic contextual information to each core user agent and assess their decision-making.

\begin{enumerate}[-]
\itemsep-0.2em
    \item Stance Alignment: We evaluate the stance of generated content, i.e., classify it into three categories: \textit{support}, \textit{neutral} and \textit{oppose}. Since the categories are concentrated on \textit{support} and \textit{neutral}, the mean absolute error (MAE) of attitude scores is also reported.
    \item Content Alignment: We classify the agent-generated content into 5 types, i.e., \textit{Call for Action}, \textit{Sharing of Opinion}, \textit{Reference to a Third Party}, \textit{Testimony} and \textit{Other} ~\cite{brunker2020role}. Accuracy and macro F1 score are reported, and cosine similarity between simulated content and real content is also provided.
    \item Behavior Alignment: We evaluate whether the agents take the corresponding actions done by users. Since only posting and retweeting can be observed in Twitter datasets, we narrow down the action space to post and retweet. Accuracy and macro F1 score are reported.
\end{enumerate}

\subsection{Macro System Evaluation}~\label{sec:4.4}
To evaluate the effectiveness of the simulation at the macro level, we quantify the attitude distribution from both horizontal and vertical perspectives in a complete multi-round simulation.
\begin{enumerate}[-]
\itemsep-0.2em
    \item Static Attitude Distribution: We capture characteristics of attitude distribution in quantitative terms: \textit{Bias} and \textit{Diversity}~\cite{lorenz2021individual}. \textit{Bias} is measured as the deviation of the mean attitude from the neutral attitude, and \textit{Diversity} is the standard deviation of attitudes. We measure at every time step and average over time. Differences between simulated and real measures $\Delta Bias$ and $\Delta Div.$ are reported.
    
    \item Time Series of the Average Attitude: We measure the similarity between the time series of average attitude and the simulated one, using Dynamic Time Warping (DTW)~\cite{muller2007dynamic} and Pearson correlation coefficient~\cite{cohen2009pearson}.
\end{enumerate}

\paragraph{Calibration and Validation}
To find the proper parameters for ABMs in the hybrid system, we perform the calibration and validation settings~\cite{gestefeld2023calibrating}. Calibration aims to find the best combination of parameters that can help match the empirical distributions. We specify parameter values for a parameter sweep to produce simulation results on E1 or P1 of each movement. Then, we report the validation results on E2 or P2. Since simulating with LLMs hundreds of times is unaffordable, we perform calibration in pure ABMs and apply the optimal parameters to the hybrid model. The details can be found in Appendix ~\ref{app:calib}.

\section{Experiments}
\subsection{Experiment Settings}
We incorporate the following ABMs for ordinary users in the hybrid framework. Meanwhile, we employ these ABMs to model all users (referred to as pure ABMs) as baselines for comparison. More details can be found in Appendix~\ref{app:abm}.
\begin{enumerate}[-]
\itemsep-0.2em
    \item Bounded Confidence Model (BC)~\cite{deffuant2000mixing}: it assumes that if the received message is close enough to an agent's attitude, the message has an assimilation force on the agent.
    \item Bounded Confidence Model-Multiple (HK)~\cite{boundedconfidence}: a variant of BC, which can handle multiple sources.
    \item Relative Agreement Model (RA)~\cite{deffuant2002can}: extends the BC  in that the similarity bias is a continuously decaying function.
    \item Social Judgement Model (SJ)~\cite{socialjudgement}: additionally includes a repulsion force based on BC.
    \item Lorenz~\cite{lorenz2021individual}: includes assimilation force, reinforcement force, similarity bias, polarization factor and source credibility.
\end{enumerate}

For all the experiments, we use GPT-3.5-Turbo-0613 to simulate core users, with max tokens set to 256 and temperature set to 0 for more deterministic results. We implement LLM-empowered core users based on AgentVerse~\cite{chen2023agentverse}, while for ordinary users we use the mesa~\footnote{https://mesa.readthedocs.io/en/stable/} library to implement the conventional agent-based models. For micro alignment evaluation, we sample (user, context) pairs from the dataset detailed in Appendix~\ref{app:micro_data} to reduce cost. For macro system evaluation, we run 14 steps for each event and more details can be found in Appendix~\ref{app:simu}.

\input{tabs/micro}

\input{tabs/macro}

\subsection{Micro Alignment Evaluation}
Table~\ref{tab:micro} shows the results of the micro alignment evaluation on the three datasets. We can observe:
\begin{enumerate}[-]
\itemsep-0.2em
    \item LLM-empowered agents effectively model core users' stances on specific topics. This can be attributed to the personalized profiles reflecting users' leanings. However, they struggle to generate non-supportive content, resulting in low F1 scores. This is because LLMs tend to produce content with clear stances, unlike real users, who may illustrate more complex behaviors, such as sharing external links or mentions.
    \item The LLM-empowered agents can replicate the content generated by users. Both in the real data and simulated results, user-generated content is concentrated in \textit{call for action} and \textit{sharing of opinions}. It's difficult for agents to generate testimony content since they lack the offline experience of the users. Moreover, the overall cosine similarity between real and simulated content approaches 80\%, affirming their capability to replicate user responses to specific contexts.
    \item The LLM-empowered agents can well distinguish between different users who are more inclined to create original content and those who prefer to retweet, achieving over 72\% accuracy on all the three datasets. It can be attributed to the portrayal of social traits and communication roles in the profile, which indirectly influences the agents' choice of actions. The ablation study in Appendix~\ref{app:ablation} further demonstrates this.
\end{enumerate}

\subsection{Macro System Evaluation}\label{sec:5.2}
Table~\ref{tab:macro} shows the results of the macro system evaluation. We can observe:
\begin{enumerate}[-]
\itemsep-0.2em
    \item Overall, the hybrid models outperform pure ABMs in terms of both static measures and time series measures. Among the models, those based on the RA and Lorenz demonstrate advantages across various datasets, benefiting from the ability of RA and Lorenz in modeling situations of extremism~\cite{chuang2023computational}.
    \item Hybrid models usually exhibit higher attitude bias compared with the corresponding pure ABMs. It's also the result of the LLMs' leaning to generate content with clear stances. With more agents modeled as positive towards the topic, the overall level of attitude is also overestimated.
    \item Taking advantage of the accurate replication of attitudes of core users empowered by LLMs, even when pure ABMs fail to capture the overall trend in attitude changes, the overall trend can be corrected in the hybrid models under the guidance of these powerful LLM-based agents.
\end{enumerate}

\subsection{Scalability Analysis}

\begin{figure}[!t]
    \centering
    \includegraphics[width=\linewidth]{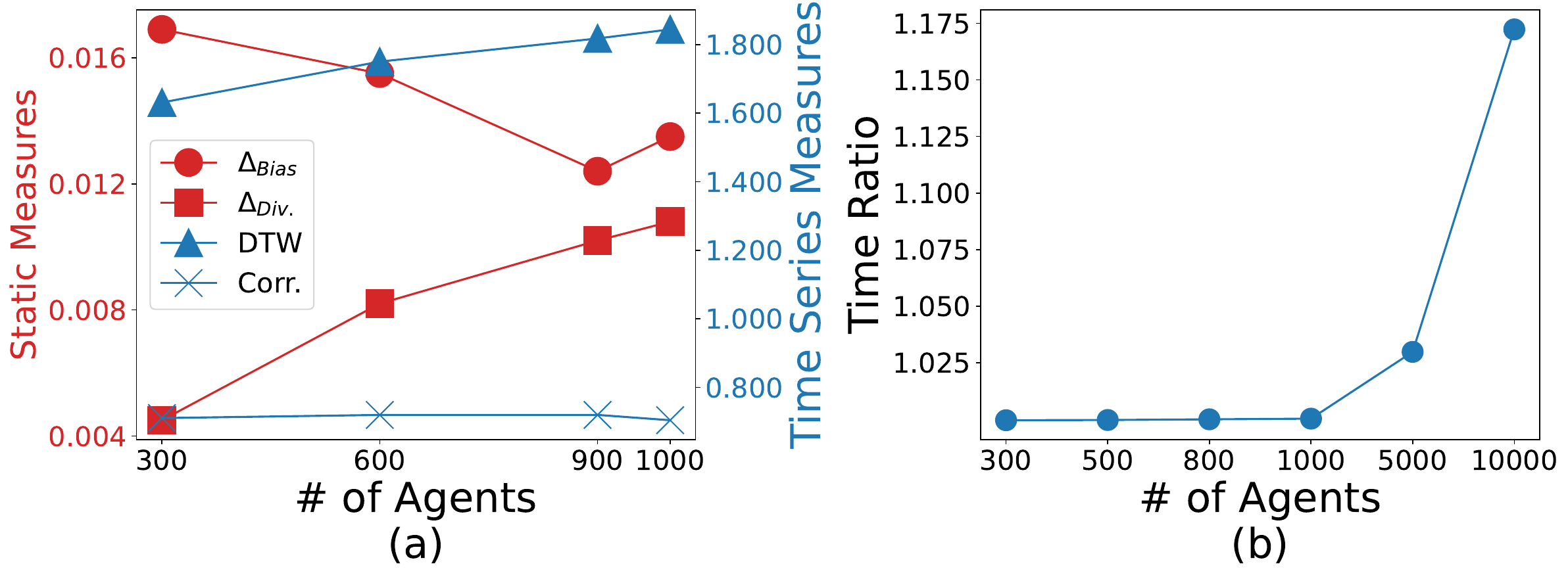}
    \caption{(a) System metrics when simulating with varying numbers of agents (better viewed in color); (b) Running Efficiency with varying numbers of agents.}
    \label{img:efficiency}
\end{figure}

Figure ~\ref{img:efficiency} depicts the performance and runtime variations observed in the \textit{Metoo} experiment across different numbers of agents. For comparative analysis, we establish the real distribution of 1,000 agents in Sec.~\ref{sec:5.2} as the reference and assess the performance of hybrid models featuring 300 core users alongside varying numbers of ordinary users. In Figure ~\ref{img:efficiency}a, except for $\Delta_{Bias}$, all other metrics exhibit only a slight decline as the proportion of ordinary users increases, indicating that a sampling simulation approach can yield competitive results. Figure ~\ref{img:efficiency}b illustrates the simulation time ratio with varying numbers of agents relative to the time in the main experiment with 1,000 agents, where the core users remain fixed at 300. Notably, the runtime primarily depends on the time required for LLM API key invocation, with scaling up the number of ordinary users hardly imposing additional burden, unless the simulation population exceeds millions, when further engineering enhancements and hardware optimizations become necessary. This observation underscores the scalability of our method.

\subsection{Further Analysis}
We further discuss more details about the simulation of the LLM-empowered core users, offering insights to enhance community communication.

\subsubsection{Replication of Echo Chambers}\label{sec:5.4.1}
\begin{figure}[!t]
    \centering
    \includegraphics[width=\linewidth]{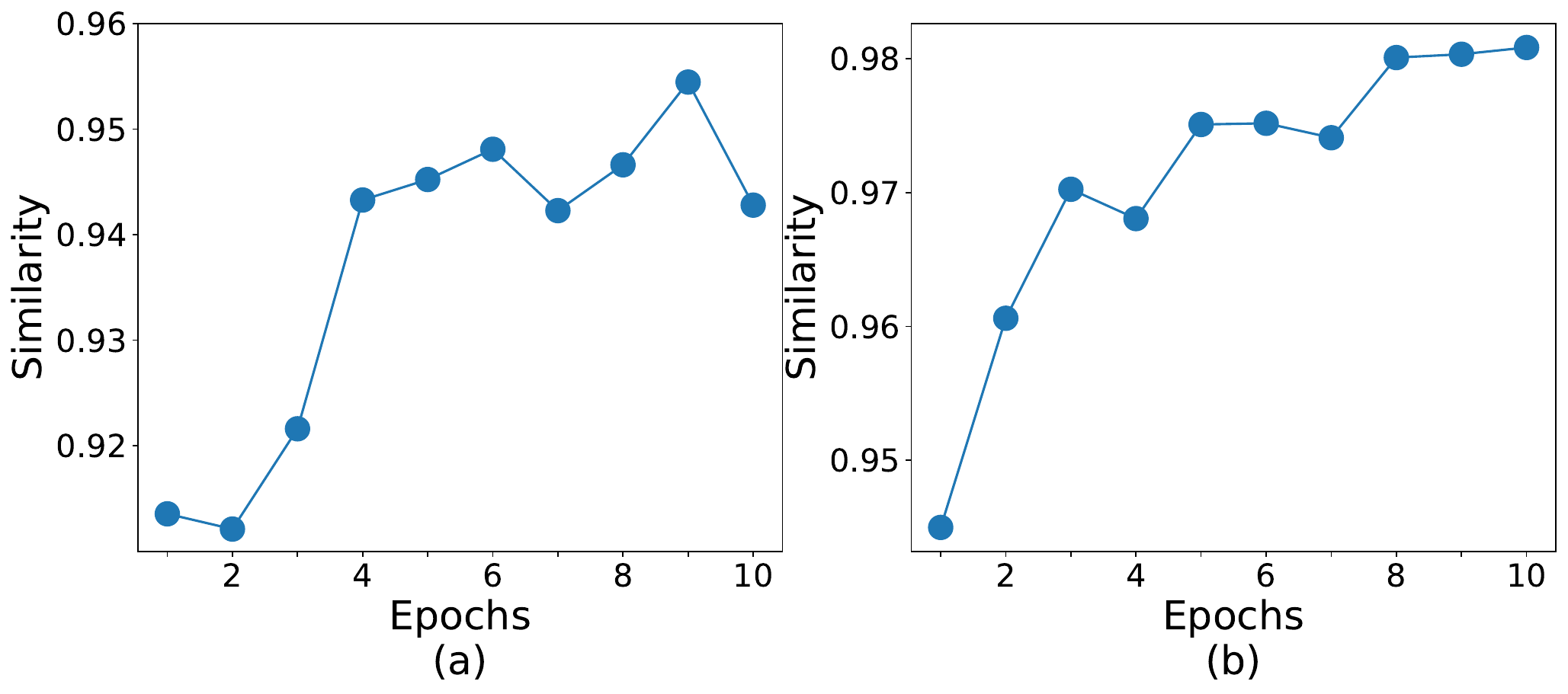}
    \caption{(a) Similarity of content of consumption and production on \textit{Metoo} simulation; (b) Similarity of content of consumption and production on \textit{Roe} simulation. \textit{BLM} is not reported since it is a partial phrase.}
    \label{img:echo}
\end{figure}

We aim to assess whether the simulation can replicate the echo chamber, a common phenomenon in online social networks. We explore this question from the perspective of the consumption and production of content~\cite{garimella2018political}. The content of production is the content generated by the agents, while the content of consumption is that generated by agents they ``follow''. The similarity between production and consumption indicates the users' tendency to consume content that is similar to their own. Figure~\ref{img:echo} reveals that as the number of epochs increases, the average similarity shows an overall upward trend. These results validate the system's capability to reflect the echo chambers.

\subsubsection{Intervention: Break the Echo Chambers}
\input{tabs/echo}

Given that echo chambers often contribute to polarization, we aim to explore strategies to mitigate this effect while safeguarding users' freedom of expression. We propose and test three solutions in our simulation framework: \textbf{S1} - Feeding the opposite opinions; \textbf{S2} - Feeding the neutral opinions; \textbf{S3} - Establishing public spaces for debate or discussion, achieved by encouraging users to share opinions using platform-provided public hashtags. We experiment on the \textit{Metoo} dataset and evaluate the homogeneity and toxicity of different situations. The homogeneity is measured by the method in Sec.~\ref{sec:5.4.1} and the toxicity is measured using the Perspective API~\footnote{\url{https://developers.perspectiveapi.com/}}.
Table~\ref{tab:echo} illustrates that all three approaches can reduce the echo chambers, but the introduction of opposing opinions can increase the toxicity of the community, while establishing spaces for open discussion can promote more peaceful exchanges.

\section{Related Work}
\subsection{Social Media User Modeling}
Social media user modeling initially focused on representing users and predicting their attributes. The earliest methods concentrate on text, using feature engineering~~\cite{schwartz2013personality, preoctiuc2018user} or deep models~\cite{huang2019hierarchical} to map user-generated content into representations. Beyond text, social networks were later incorporated into user modeling, either through an explicit combination of textual and graph encoders~\cite{xiao2020timme, mou2021align} or through implicit pre-training patterns~\cite{jiang2023retweet, wu2024pasum}. However, these methods were limited to predicting discrete attributes or behaviors and could not predict more complex behaviors such as specific content generation. The development of LLMs has laid the foundation for simulating such complex scenarios.

\subsection{LLM-empowered Autonomous Agents}
With the prominent development of Large Language Models, the LLM-empowered autonomous agent has recently gained significant attention. Integrating profile, memory, reflection and planning modules, ~\citeauthor{park2023generative} design generative agents to simulate the human daily life. Based on this universal framework, agents with slightly different architectures have been widely applied in various scenarios and applications. Among these works, some are utilized for task-solving purposes, i.e., executing pre-defined tasks, such as software development~\cite{qian2023communicative, qian2023experiential,hong2023metagpt}, collaboration~\cite{chen2023agentverse}, and exploring the world in Minecraft~\cite{wang2023voyager}. Others focus on simulation, where the replication of human behaviors is focused. Scenarios including social interaction,~\cite{park2022social,liu2023training} recommendation~\cite{wang2023recagent, zhang2023generative}, the world war~\cite{hua2023war} and communication game~\cite{xu2023exploring} have been explored to provide insights on communication. Most works only require a small number of agents and haven't considered scenarios requiring large-scale agents.

\subsection{Social Simulation}
Agent-based models for social simulation on opinion dynamics mainly focus on how individuals change their attitudes due to others' influence~\cite{chuang2023computational, chuang2023simulating}. These models can be divided into deductive ABMs and inductive ABMs. Represented by the Bounded Confidence Model~\cite{deffuant2000mixing}, the former category is based on psychology, such as the social judgment theory~\cite{sherif1961social}. By contrast, the latter often involves human experiments, which are expensive and with limited scale. With ability in human-level intelligence, LLM-empowered agents have potential to serve as a substitute for human subjects. \citeauthor{park2022social} and \citeauthor{tornberg2023simulating} provide a simulation platform to help designers see beyond social interactions that people intend and improve social interaction. Sotopia~\cite{zhou2023sotopia} designs an evaluation framework for social intelligence. $S^3$~\cite{gao2023s} simulate public opinion through Markov Chain and LLMs, but how they deal with the large scale of users remains ambiguous.

\section{Conclusion}
In this paper, we propose a hybrid framework for social media user simulation. We empower core users and ordinary users with LLMs and ABMs, and we provide a Twitter-like environment and a benchmark SoMoSiMu-Bench for simulation and evaluation. Experiment results demonstrate the effectiveness and flexibility of our method.

\section*{Acknowledgement}
This work is supported by National Natural Science Foundation of China (No. 62176058) and National Key R\&D Program of China (2023YFF1204800). The project's computational resources are supported by CFFF platform of Fudan University and the data collecting is supported by DBCloud. We would also like to thank the chairs and reviewers for their constructive feedback.

\section*{Limitations}
Our work is the first step towards a large-scale simulation implemented by a hybrid framework and it is limited in two aspects.  In terms of data, although we have incorporated a larger number of agents than other studies, due to limitations in annotation costs, we have not yet validated interactions among millions of agents. In terms of LLM-empowered agents, due to reinforcement learning techniques, LLMs are also biased toward being more polite, articulate and respectful than users on real-world social media platforms~\cite{tornberg2023simulating}, bringing bias to our study. More careful prompt engineering will be considered to solve this problem in future work.

\section*{Ethics Statement}
\paragraph{Data Collection and Privacy}
Our data collection is in compliance with Twitter’s terms of service and matches previous publications. Although tweets are public, when releasing data, we will share tweet id rather than raw data, to minimize the privacy risk. Furthermore, during the simulation, we anonymize each user by renaming them.
\paragraph{Simulation for Social Good}
The purpose of this paper is to use simulation to recreate the real situation of social movements and provide insights for improving harmonious communication among users. But it might also be misused to label people with a specific label that they do not want to be associated with. We suggest that when in use the tools should be accompanied by descriptions about their limitations and imperfect performance, as well as allow users to opt out from being the subjects of measurement.

\bibliography{acl2023}
\bibliographystyle{acl_natbib}

\newpage
\appendix
\input{apps/abm}

\input{apps/data}

\input{apps/simu}
\input{apps/res}

\end{document}

%% file: tabs/data.tex
\begin{table}[!t]
\centering
\resizebox{0.48\textwidth}{!}{
\begin{tabular}{@{}llllr@{}}
\toprule
\textbf{Dataset}                & \textbf{Event} & \textbf{\#Users} & \textbf{\#Tweets} & \multicolumn{1}{l}{\textbf{Time Span}} \\ \midrule
\multirow{2}{*}{Metoo} & E1    & 1,000   & 18,638   & Oct 15 - Oct 22, 2017         \\
                       & E2    & 1,000   & 13,291   & Jan 06 - Jan 13, 2018         \\
\multirow{2}{*}{Roe}   & E1    & 1,000   & 61,687   & May 02 - May 09, 2022         \\
                       & E2    & 1,000   & 59,829   & Jun 24 - Jul 01, 2022         \\
\multirow{2}{*}{BLM}   & P1    & 1,000   & 10,710   & May 25 - Jun 01, 2020         \\
                       & P2    & 1,000   & 21,480   & Jun 02 - Jun 09, 2020         \\ \bottomrule
\end{tabular}}
\caption{Statistics of our dataset. In \textit{Metoo},  E1 is \textit{American actress Alyssa Milano starts the \#Metoo movement} and E2 is \textit{\#Timesup campaign on the 2019 Golden Globes Awards}; In \textit{Roe}, E1 is \textit{The leakage of the Supreme Court draft opinion} and E2 is \textit{The Supreme Court overturns Roe v. Wade}; In \textit{BLM}, we include two phases after the \textit{Murder of George Floyd}.}
\label{tab:dataset}
\end{table}

%% file: tabs/micro.tex
\begin{table}[!t]
\resizebox{0.48\textwidth}{!}{
\begin{tabular}{@{}l|cccccccc@{}}
\toprule
\multirow{2}{*}{\textbf{Datasets}} & \multicolumn{3}{c}{\textbf{Stance}} & \multicolumn{3}{c}{\textbf{Content}} & \multicolumn{2}{c}{\textbf{Behavior}} \\
                                   & Acc.       & F1         & MAE       & Acc.       & F1         & Sim.       & Acc.              & F1                \\ \midrule
Metoo                              & 0.9679     & 0.3400     & 0.2311    & 0.7010     & 0.1988     & 0.8064     & 0.7313            & 0.5212            \\
Roe                                & 0.9430     & 0.3361     & 0.2058    & 0.6423     & 0.1957     & 0.8090     & 0.6665            & 0.4691            \\
BLM                                & 0.8991     & 0.3735     & 0.1627    & 0.7353     & 0.2218     & 0.8406     & 0.7796            & 0.5759            \\ \bottomrule
\end{tabular}
}
\caption{Results of micro alignment evaluation. }
\label{tab:micro}
\end{table}

%% file: tabs/macro.tex
\begin{table*}[!t]
\resizebox{\textwidth}{!}{
\begin{tabular}{@{}l|rrrr|rrrr|rrrr@{}}
\toprule
\multicolumn{1}{c|}{\multirow{2}{*}{\textbf{Method}}} & \multicolumn{4}{c|}{\textbf{Metoo}}                                                                                                  & \multicolumn{4}{c|}{\textbf{Roe}}                                                                                                    & \multicolumn{4}{c}{\textbf{BLM}}                                                                                                    \\
\multicolumn{1}{c|}{}                                 & \multicolumn{1}{l}{$\Delta_{Bias}$↓} & \multicolumn{1}{l}{$\Delta_{Div.}$↓} & \multicolumn{1}{l}{DTW↓} & \multicolumn{1}{l|}{Corr.↑} & \multicolumn{1}{l}{$\Delta_{Bias}$↓} & \multicolumn{1}{l}{$\Delta_{Div.}$↓} & \multicolumn{1}{l}{DTW↓} & \multicolumn{1}{l|}{Corr.↑} & \multicolumn{1}{l}{$\Delta_{Bias}$↓} & \multicolumn{1}{l}{$\Delta_{Div.}$↓} & \multicolumn{1}{l}{DTW↓} & \multicolumn{1}{l}{Corr.↑} \\ \midrule
BC                                           & 0.0124                             & 0.0184                             & 2.7760                   & 0.4831                     & 0.0265                             & 0.0144                             & 5.7662                   & -0.7755                    & 0.0078                             & 0.0036                             & 5.2289                   & -0.4404                    \\
Hybrid w/ BC                                 & \underline{0.0135}                             & 0.0108                             & 1.8440                   & 0.7043                     & 0.0239                             & 0.0121                             & 2.4611                   & 0.3607                     & \underline{0.0300}                             & \underline{0.0069}                             & 3.9254                   & 0.1248                     \\
HK                                           & 0.0093                             & 0.0105                             & 2.9171                   & 0.0262                     & 0.0258                             & 0.0185                             & 7.7254                   & -0.7532                    & 0.0081                             & 0.0101                             & 4.1204                   & -0.3026                    \\
Hybrid w/ HK                                 & \underline{0.0126}                             & 0.0037                             & 1.9136                   & 0.6517                     & \underline{0.0319}                             & 0.0157                             & 3.6752                   & -0.0807                    & \underline{0.0578}                             & 0.0093                             & 3.7288                   & -0.2433                    \\
RA                                           & 0.0062                             & 0.0055                             & 3.1063                   & -0.0687                    & 0.0237                             & 0.0120                             & 2.9521                   & 0.0811                     & 0.0039                             & \textbf{0.0017}                             & 3.0441                   & 0.2666                     \\
Hybrid w/ RA                                 & \underline{0.0117}                             & \textbf{0.0008}                             & \textbf{1.7829}                   & \textbf{0.7238}                     & 0.0221                             & 0.0104                             & 2.3326                   & 0.4274                     & \underline{0.0376}                             & {0.0070}                             & \textbf{2.2353}                   & \textbf{0.6050}                     \\
SJ                                           & 0.0064                             & 0.0192                             & 2.2994                   & 0.2009                     & 0.0209                             & 0.0106                             & 1.2739                   & 0.6177                     & 0.0411                             & 0.0072                             & 2.7778                   & 0.4475                     \\
Hybrid w/ SJ                                 & \underline{0.0098}                             & 0.0119                             & 2.2789                   & 0.6327                     & 0.0203                             & \textbf{0.0095}                             & 1.1896                   & 0.6598                     & 0.0076                             & 0.0018                            & 2.4564                   & 0.5167                     \\
Lorenz                                       & 0.0131                             & 0.0198                             & 5.3049                   & -0.4657                    & 0.0352                             & 0.0172                             & 1.1027                   & 0.7329                     & 0.0895                             & 0.0094                             & 2.8897                   & 0.4387                     \\
Hybrid w/ Lorenz                             & \textbf{0.0035}                            & 0.0116                             & 2.9857                   & 0.6103                     & \textbf{0.0093}                             & 0.0147                             & \textbf{1.0148}                   & \textbf{0.7576}                     & \textbf{0.0023}                             & 0.0079                             & 2.5394                   & 0.5055                     \\ \bottomrule
\end{tabular}}
\caption{Results of macro system evaluation. The average results of 3 runs are reported. In the vast majority of cases, hybrid systems show improvements across various aspects compared to ABMs. \textbf{Bold} presents the best performance in the column. \underline{Underline} indicates the metric for the hybrid model did not surpass that of the corresponding ABM.}
\label{tab:macro}
\end{table*}

%% file: tabs/echo.tex
\begin{table}[!t]
\centering
\resizebox{0.43\textwidth}{!}{
\begin{tabular}{@{}ccc@{}}
\toprule
\textbf{Method} & \textbf{Avg. Homogeneity} & \textbf{Avg. Toxity} \\ \midrule
S1              & \textbf{0.8551}                    &    0.1426                  \\
S2              & 0.8580                    &    0.1296                  \\
S3              & 0.8962                    &    \textbf{0.1163}                  \\ \bottomrule
\end{tabular}
}
\caption{Results of solutions to break the echo chambers. \textbf{Bold} presents the best performance in the column.}
\label{tab:echo}
\end{table}

%% file: apps/abm.tex
\section{Agent-based Models}~\label{app:abm}
In this section, we present the agent-based models based on different psychological theories in detail, following the unified formulation proposed by ~\citet{chuang2023computational}. Assume that there are $N$ agents with index $i\in \{1,2,...,N\}$ in a system $I_{system}$, where each agent has an attitude toward the same topic. Let $a_{i,t}$ be the attitude score of agent $i$ at time $t$, $J_{i, t}$ be the set of agents that will influence agent $i$ at time $t$, $m_{j,t}$ be the message that agent $j$ convey to other agents at time $t$.

\subsection{Bounded Confidence Model (BC)}
The bounded-confidence (BC) model was proposed by ~\cite{deffuant2000mixing}. It assumes that when the message $m_{j,t}$ is close enough to the agent $i$'s attitude $a_{i,t}$, the message exerts an assimilation force on the agent's attitude.

\paragraph{Update Function} The attitude update function is:
\begin{align}
    \Delta a_{i, t}=\alpha_i \cdot \operatorname{sim}\left(a_{i, t}, m_{j, t}\right) \cdot\left(a_{j, t}-a_{i, t}\right),
\end{align}
where
\begin{align}
    \operatorname{sim}\left(a_{i, t}, m_{j, t}\right)=\left\{\begin{array}{c}
1, \text { if }\left|a_{j, t}-a_{i, t}\right|<\varepsilon_i \\
0, \text { otherwise }
\end{array} ,\right.
\end{align}

\paragraph{Selection Function} The selection function is:
\begin{align}
    \begin{split}
    J_{i, t}= & f_{\text {select }}(i, t)=\{ \text{one random agent } j  \\
    & \text{ in the system within the confidence bound,} \\
    & \text{ i.e., whose} \left|a_{j, t}-a_{i, t}\right|<\varepsilon_i,\text{ except the }\\
    & \text{agent } i \text{ itself} \},
    \end{split}
\end{align}
 where the number of source agents is set to $N_J=1$.

\paragraph{Message Function} The message function is:
\begin{align}
    m_{j, t}=f_{\text {message }}\left(a_{j, t}\right)=a_{j, t},
\end{align}

\paragraph{Other Assumptions}
$\alpha_i \in [0, 1]$

\subsection{Bounded Confidence Model-Multiple (HK)}
\citet{boundedconfidence}'s bounded confidence model is an extension of the bounded confidence model, where it can handle multiple source agents, i.e., $1<=N_J<=N$. 

\paragraph{Update Function} The update function is:
\begin{align}
    \begin{split}
    \Delta a_{i, t}= & \frac{N_{\varepsilon_i}}{\left(N_{\varepsilon_i}+1\right)} \cdot \frac{1}{\left(N_{\varepsilon_i}\right)} \\
    & \cdot \sum_{j \in J_{i, t}} \operatorname{sim}\left(a_{i, t}, a_{j, t}\right) \cdot\left(a_{i, t}-a_{j, t}\right),
    \end{split}
\end{align}
where 
\begin{align}
    \operatorname{sim}\left(a_{i, t}, m_{j, t}\right)=\left\{\begin{array}{c}
1, \text { if }\left|a_{j, t}-a_{i, t}\right|<\varepsilon_i \\
0, \text { otherwise }
\end{array} ,\right.
\end{align}

$\varepsilon_i$ is the confidence bound, $N_{\varepsilon_i}$ is the number of agents within the confidence bound.

\paragraph{Selection Function} The selection function is:
\begin{align}
    \begin{split}
    J_{i, t}= &f_{\text {select }}(i, t)=I_{\text {system }} \backslash\{i\} = \text { \{all the } N \\
    & \text { agents in the system except the agent } i \},
    \end{split}
\end{align}
Note that only those within the confidence bound influence the agent $i$'s attitude.

\paragraph{Message Function}The message function is:
\begin{align}
    m_{j, t}=f_{\text {message }}\left(a_{j, t}\right)=a_{j, t},
\end{align}

\paragraph{Other Assumptions}
The strength of the social influence is a function of the number of agents in the confidence bound, i.e., $\alpha_i\left(N_{\varepsilon_i}\right)=\frac{N_{\varepsilon_i}}{\left(N_{\varepsilon_i}+1\right)}$.

\subsection{Relative Agreement Model (RA)}
The relative agreement (RA) model extends the bounded confidence model in that the similarity bias $asm(a_{i,t}, m_{j,t})$ is a continuously decaying function. The RA model assumes that the more the message $m_{j,t}$ agrees with $a_{i,t}$, the more susceptible the agent $i$ is to its assimilation force.

In the RA model, there is an uncertainty $u_{i,t}>0$ around each agent's attitude $a_{i,t}$. Therefore, the agent $i$'s attitude is modeled as a segment $seg_{i,t}=[a_{i,t}-u_{i,t}, a_{i,t}+u_{i,t}]$. When agent $j$ influences the agent $i$, the larger their segments "agree", the larger the influence should be.

\paragraph{Update Function} The update function is:
\begin{align}
    \Delta a_{i, t}=\alpha_i \cdot \operatorname{sim}\left(a_{i, t}, u_{i, t}, a_{j, t}, u_{j, t}\right) \cdot\left(a_{i, t}-a_{j, t}\right),
\end{align}
where
\begin{align}
    \operatorname{sim}\left(a_{i, t}, u_{i, t}, a_{j, t}, u_{j, t}\right)=\left\{\begin{array}{c}
\left(h_{i, j} / u_j\right)-1, \\
 \text { if }\left(h_{i, j} / u_j\right)>1 \\
0, \text { otherwise }
\end{array} ,\right.
\end{align}

$h_{i,j}$ is the overlap between $seg_{i,t}$ and $seg_{j,t}$. The term $h_{i,j}/u_j$ is referred to as the 
relative agreement of the agent $j$ with the agent $i$.

\paragraph{Selection Function} The selection function is:
\begin{align}
    \begin{split}
    J_{i, t}=&f_{\text {select }}(i, t)=\{\text { one random agent } j \text { in }  \\
    & I_{\text {system }} \text {, except the agent } i \text{ itself\}},
    \end{split}
\end{align}

\paragraph{Message Function} The message function is:
\begin{align}
    \begin{split}
    m_{j, t}&=f_{\text {message }}\left(\left[a_{j, t}-u_{j, t}, a_{j, t}+u_{j, t}\right]\right)\\
    & =\left[a_{j, t}-u_{j, t}, a_{j, t}+u_{j, t}\right],
    \end{split}
\end{align}

\paragraph{Other Assumptions}
$\alpha_i \in [0,1]$.

\subsection{Social Judgement Model (SJ)}
The social judgment (SJ) model differs from the bounded confidence model in that it additionally includes a repulsion force. It assumes that an agent will assimilate towards the message $m_{j,t}$ if it is close enough to its attitude $a_{i,t}$, and will distance away from the message if it is too far away from its attitude.

\paragraph{Update Function} The attitude update function is:
\begin{align}
    \begin{split}
    \Delta a_{i, t}=&\alpha_i \cdot[\operatorname{sim}\left(a_{i, t}, m_{j, t}\right) \cdot\left(a_{j, t}-a_{i, t}\right)\\
    & +\operatorname{rep}\left(a_{i, t}, m_{j, t}\right)],
    \end{split}
\end{align}
where 
\begin{align}
    \begin{split}
    &\operatorname{sim}\left(a_{i, t}, a_{j, t}\right) \cdot\left(a_{j, t}-a_{i, t}\right)=\\
    &\left\{\begin{array}{c}
\left(a_{j, t}-a_{i, t}\right), \\
\text { if }\left|a_{j, t}-a_{i, t}\right|<u_i \\
0 \text {, otherwise }
\end{array}\right.,
    \end{split}
\end{align}
\begin{align}
    \operatorname{rep}\left(a_{i, t}, a_{j, t}\right)=\left\{\begin{array}{c}
-\left(a_{j, t}-a_{i, t}\right),\\
    \text { if }\left|a_{j, t}-a_{i, t}\right|>t_i \\
0, \text { otherwise }
\end{array}\right.,
\end{align}

$u_{i}$ and $t_i$  specify the latitude of acceptance and the latitude of rejection, respectively.

\paragraph{Selection Function} The selection function is:
\begin{align}
    \begin{split}
    J_{i, t}=&f_{\text {select }}(i, t)=\{\text { one random agent } j \text { in } \\
    & I_{\text {system }} \text {, except the agent } i \text{ itself\}},
    \end{split}
\end{align}

\paragraph{Message Function} The message function is:
\begin{align}
    m_{j, t}=f_{\text {message }}\left(a_{j, t}\right)=a_{j, t},
\end{align}

\paragraph{Other Assumptions} $\alpha_i \in [0,1]$.

\begin{figure*}[!t]
    \centering
    \includegraphics[width= \linewidth]{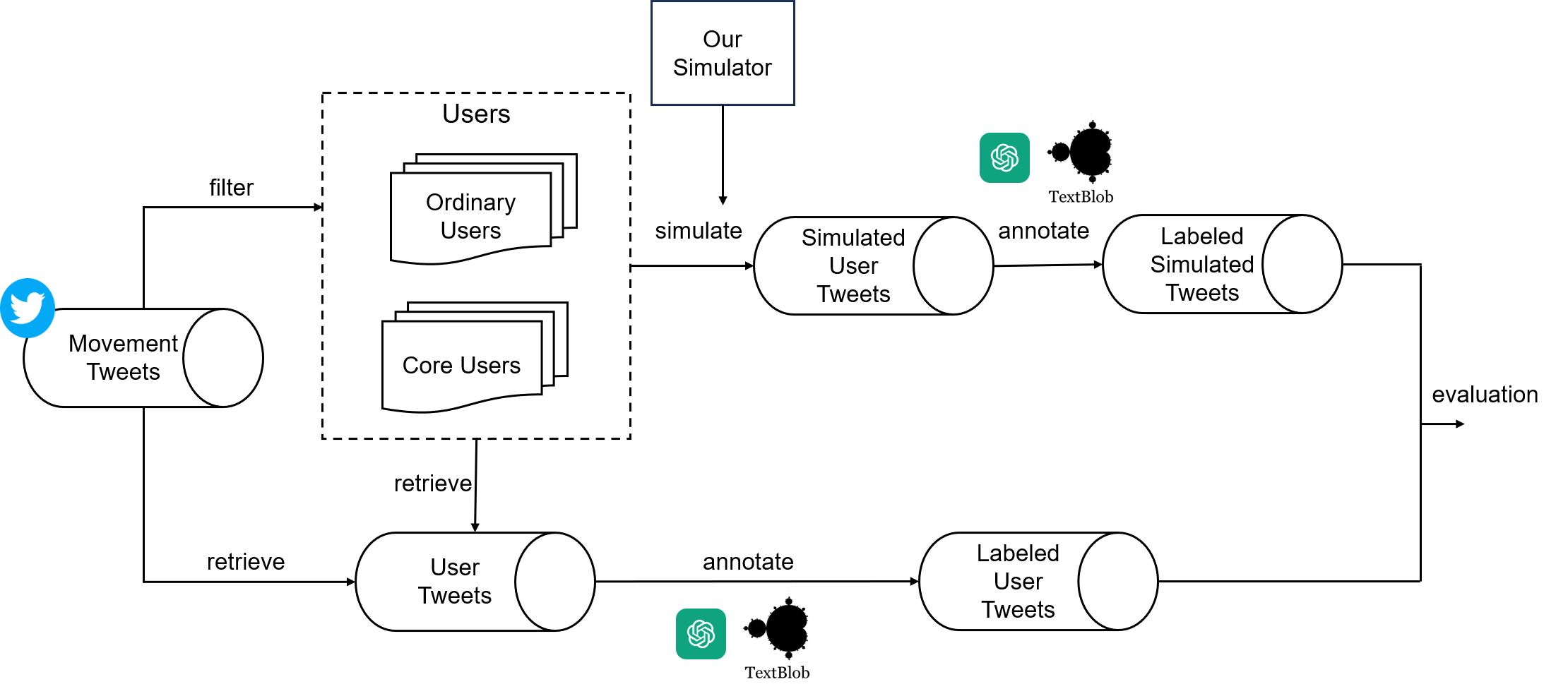}
    \caption{The data processing process.}
    \label{img:data_process}
\end{figure*}

\subsection{Lorenz Model}
\citet{lorenz2021individual} proposed a model which includes assimilation force, reinforcement 
force, similarity bias, polarization factor, and source credibility. 

\paragraph{Update Function} The attitude update function is:
\begin{align}
    \begin{split}
\Delta a_{i, t}=&\alpha_i s(i, j) p o l\left(a_{i, t}\right) \operatorname{sim}\left(a_{i, t}, m_{j, t}\right) \\
&\cdot\left[\rho \cdot \operatorname{asm}\left(a_{i, t}, m_{j, t}\right)+(1-\rho) \cdot \operatorname{ref}\left(m_{j, t}\right)\right],
\end{split}
\end{align}
where the polarization factor is:
\begin{align}
    \operatorname{pol}\left(a_{i, t}\right)=\frac{M^2-a_{i, t}{ }^2}{M^2},
\end{align}
The hyperparameter parameter $M$ is the theoretical boundary for the attitude space.
The similarity bias is:
\begin{align}
    \operatorname{sim}\left(a_{i, t}, m_{j, t}\right)=\frac{\lambda^k}{\lambda^k+\left|m_{j, t}-a_{i, t}\right|^k},
\end{align}
The hyperparameters $\lambda$ and $k$ specify the shape of the similarity bias function.
The reinforcement force is:
\begin{align}
    ref(m_{j,t})=m_{j,t},
\end{align}
The assimilation force is:
\begin{align}
    asm(a_{i,t}, m_{j,t}) = (m_{j,t}-a_{i,t}),
\end{align}

\paragraph{Selection Function}  The selection function is:
\begin{align}
    \begin{split}
    J_{i, t}=&f_{\text {select }}(i, t)=\left\{\text { one random agent } j \text { in } I_{\text {system }}\right. \\
    & \text {, except the agent } i \text{ itself\}},
    \end{split}
\end{align}

\paragraph{Message Function} The message function is:
\begin{align}
    m_{j, t}=f_{\text {message }}\left(a_{j, t}\right)=a_{j, t},
\end{align}

%% file: apps/data.tex
\section{Data Collection}~\label{app:data}

\subsection{Tweets of Social Movements}
We get tweet ids of the three social movements from ~\citeauthor{metoodata}, ~\citeauthor{chang2023roeoverturned} and ~\cite{blmdata}, and crawl those during the required event periods. Overall, a total of 4,985,000, 22,869,406, and 25,112,678 tweets were collected. Subsequently, we sample users described in Sec.~\ref{sec:4.1} from the authors of these tweets. After collecting these data, the post-processing process is shown in Figure ~\ref{img:data_process}.

\subsection{Profile Construction}\label{app:profile}
As mentioned in Sec.~\ref{sec:profile}, we include demographics, social traits and communication roles in the profile of each agent. We construct profiles from the bios and tweets of Twitter users.
\paragraph{Demographics} In order to minimize the noise caused by annotating, we first identify gender, political leaning and account type from one's bio using regular expression with strict rules. Then, we prompt GPT-3.5-Turbo-0613 to infer those can not be directly matched. The candidate list for account type was acquired from~\cite{brunker2020role}: [\emph{Journalist, Private Person, Celebrity, Media Organization, Activist, Politician, Social Bot, NGO, International Organization, Company, Governmental Organization, Suspended Accounts}].

\paragraph{Social Traits}
The activity level and influence level are measured by the number of tweets and number of followers for each user. We follow ~\cite{zhang2023generative} to allocate the social trait of activity and influence to each user based on the ascending order of the measurement with a ratio of 6:3:1. The activity levels include [\emph{not active, moderately active, highly activity}] and the influence levels include [\emph{not influential, moderately influential, highly influential}].

\paragraph{Communication Role} We identify the communication roles of each user by calculating the metrics proposed by ~\cite{tinati2012identifying} and ~\cite{bentwooddistributed}. Then, we assign corresponding descriptions of the roles when generating profiles of users. 
\begin{itemize}
    \item Idea Starter: Start a conversational meme, and tend to be highly engaged with the media and post original content.
    \item Amplifier: Collect multiple thoughts and share ideas and opinions. Enjoy being the first one to retweet original content.
    \item Curator: Use a broader context to define ideas. Tend to take ideas of others and either validate, question, challenge, or dismiss them. Tend to be the ties that form between others, aggregating ideas together to help clarify and steer the topic of conversation.
    \item Commentator: Detail and refine ideas. Take part in something to which he or she strongly feels about. Want to share information not for self-benefit.
    \item Viewer: Take a passive interest in the conversation. Leave footprint by viewing rather than contributing to the conversation. Prefer to consume information rather than create or share information online.
\end{itemize}

After acquiring the attributes of the user, we prompt GPT-3.5-Turbo-0613 to rephrase the profile using natural languages. The prompt template is as follows:
\begin{center}
\begin{tcolorbox}[colback=red!5!white,colframe=red!55!black,width=0.95\columnwidth,title={Prompt for Profile Generation}]
Given the following observation about an individual \{name\}, please summarize the relevant details from the profile. His or her profile information is as follows:\\
\\
Name: \{name\}\\
Gender: \{gender\}\\
Political Leaning: \{ideo\}\\
Activity Level: \{activity\}\\
Influence Level: \{influence\}\\
Feature: \{commu\_role\}\\
Account Type: \{account\_type\}\\
Short Bio: \{bio\}\\
A selection of posted tweets: \{tweets\}\\
You can deduce the preferences and personality from the bio and tweets, but please avoid repeating the observation in the summary.\\
Summary:
\end{tcolorbox}
\label{tab:prompt_progile}
\end{center}

\subsection{Annotation of User-generated Content}
We annotate the stance and content type expressed in users' tweets and simulated content using GPT-3.5-Turbo-0613 with temperature set to 0:

\begin{center}
\begin{tcolorbox}[colback=red!5!white,colframe=red!55!black,width=0.95\columnwidth,title={Prompt for Content Annotation}]
Please classify the text into one of the following categories based on its content. Only output your choice.\\
\\
1. call for action: tweet contained a call for action (e.g. requesting, challenging, promoting, inviting, summoning someone to do something).\\
2. testimony: tweet contained a testimony of the victim (e.g. report, declaration, first-person experience).\\
3. sharing of opinion: e.g. evaluation, appreciation, addition, analysis of opinions.\\
4. reference to a third party: reporting on something/-one, direct and indirect quotes.\\
5. other: other content that does not fall into the above categories.\\
\\
Text: \{text\} \\
Answer: 
\end{tcolorbox}
\label{tab:prompt_content}
\end{center}

\begin{center}
\begin{tcolorbox}[colback=red!5!white,colframe=red!55!black,width=0.95\columnwidth,title={Prompt for Stance Annotation}]
What's the author's stance on \{target\}? Please choose from Support, Neutral, and Oppose. Only output your choice.\\
\\
Text: \{text\} \\
Stance: 
\end{tcolorbox}
\label{tab:prompt_stance}
\end{center}

\subsection{Micro-level Dataset Description}~\label{app:micro_data}
\begin{table}[!t]
\resizebox{0.48\textwidth}{!}{
\begin{tabular}{@{}lrrrr@{}}
\toprule
\multicolumn{1}{c}{\textbf{Dataset}} & \multicolumn{1}{c}{\textbf{Size}} & \multicolumn{1}{c}{\textbf{Stance}} & \multicolumn{1}{c}{\textbf{Content}} & \multicolumn{1}{c}{\textbf{Behavior}} \\ \midrule
Metoo                                & 2,214                             & 2,166:33:15                         & 89:78:67:1,792:188                   & 422:1,792                             \\
Roe                                  & 3,595                             & 3,528:29:38                         & 48:6:72:3,362:107                    & 233:3,362                             \\
BLM                                  & 971                               & 934:33:4                            & 31:10:20:887:23                      & 84:887                                \\ \bottomrule
\end{tabular}}
\caption{Description of Dataset for Micro-level Alignment Evaluation. \emph{Stance} column indicates the distribution of support, neutral and oppose; \emph{Content} column indicates the distribution of call for action, testimony, sharing of opinion, reference to a third party and other; \emph{Behave} column indicates the distribution of post and retweet.}
\label{tab:micro_dataset}
\end{table}

When preparing micro-level (user, context) samples, we randomly sample from the datasets, instead of using the whole dataset to reduce cost. When sampling, we set some rules to remove those difficult or ambiguous for LLMs to annotate: (1) We remove samples whose ground truth response tweets have fewer than 10 words after removing hashtags and URLs. (2) We remove tweets that are merely direct reposts of news articles without expressing their opinions. As a result, we got datasets for micro-level alignment shown in Table~\ref{tab:micro_dataset}. We have manually reviewed 100 random samples for each label category and found the GPT annotation consistency to be 0.93 and 0.92 for stance and content respectively.

%% file: apps/simu.tex
\section{Simulation Details}~\label{app:simu}
\subsection{Simulation Process}~\label{app:alg}
The simulation processes for micro-level user behavior replication and macro-level opinion dynamics modeling are shown in Algorithm~\ref{alg:single} and Algorithm~\ref{alg:multi} respectively.
\begin{algorithm}
\caption{Single-round Simulation for Core User Behavior Replication}
\begin{algorithmic}[1]
    \STATE \textbf{Inputs:} Core users $U$ and corresponding real contexts $C$, size of (user, context) pairs $N$
    \STATE \textbf{Outputs:} Predicted behavior of each agent for users $U$
    \STATE \textbf{Initialize agents:}
    \FOR{$i$ in $1$ to $N$}
        \STATE Assign the profile of user $u_i$ to its agent
    \ENDFOR
    \STATE \textbf{Simulate:}
    \FOR{$i$ in $1$ to $N$}
        \STATE agent $u_i$ generates response based on its profile and context $c_i$
    \ENDFOR    
    \RETURN simulated response of each (user, context) pair
\end{algorithmic}
\label{alg:single}
\end{algorithm}

\begin{algorithm}
\caption{Multi-round Simulation for Opinion Dynamics Forecasting}
\begin{algorithmic}[1]
    \STATE \textbf{Inputs:} Core users $U^c$ and ordinary users $U^o$
    \STATE \textbf{Outputs:} Attitude scores for each agent of users in $U^c$ and $U^o$, at time $1$ to $T$
    \STATE \textbf{Initialize agents:}
    \FOR{each user agent $i$ in $1$ to $U^c$}
        \STATE Assign the profile of user $u^c_i$ to its agent
        \STATE Set initial attitude score $a_{i,t}$
        \STATE Set visible agent set according to the authentic social networks
    \ENDFOR
    \FOR{each user agent $j$ in $1$ to $U^o$}
        \STATE Set initial attitude score $a_{j,t}$
    \ENDFOR
    \STATE \textbf{Simulate continuous interactions:}
    \FOR{each time step in $1$ to $T$}
        \FOR{each user agent $i$ in $1$ to $U^c$}
            \STATE Manipulate the memory by reflecting and retrieving relevant memory
            \STATE Generate response $r_{i,t}$ based on its profile, memory, triggering news, Twitter page and notifications at time $t$
            \STATE Update the attitude score $a_{i,t}$ according to $r_{i,t}$
            \STATE Manipulate the memory by writing observations
        \ENDFOR
        \FOR{each user agent $j$ in $1$ to $U^o$}
            \STATE Select the set of agents to interact with $A_{j,t}$ through the selection function
            \STATE Update the attitude score $a_{j,t}$ based on the update function and attitude scores of agents in $A_{j,t}$
        \ENDFOR
    \ENDFOR
    \RETURN Attitude scores for each agent of users in $U^c$ and $U^o$, at time $1$ to $T$
    
\end{algorithmic}
\label{alg:multi}
\end{algorithm}

\subsection{Simulation Settings}
For all the experiments, we use GPT-3.5-Turbo-0613 to simulate core users, with max tokens set to 256 and temperature set to 0 for more deterministic results. We run the simulator on a Linux server with a single NVIDIA GeForce RTX 4090 24GB GPU and an Intel(R) Xeon(R) Gold 6226R CPU. The average cost for the whole simulation for an event is around 20-30 dollars. (Note that the cost was estimated when GPT-3.5-Turbo points to GPT-3.5-Turbo-0613, so the costs may increase now since GPT-3.5-Turbo-0613 becomes one of the older models). For each event, we simulate 14 steps, with the step size estimated by calculating the average posting interval of users in the empirical data. For the validation of pure ABMs, we report the average results of 10 runs. For the validation of hybrid models, we simulate once and keep the generated content of core use agents to run the hybrid systems 10 times and report the average results, instead of directly running the hybrid system 10 times to reduce costs.

\subsection{Prompt and Response Example for Core User Agent}~\label{app:prompt4user}
\begin{center}
\begin{tcolorbox}[breakable, colback=red!5!white,colframe=red!55!black,width=0.95\columnwidth,title={Prompt Example of Core Users in Metoo Movement.}]
You are using the social media Twitter. You might need to perform reaction to the observation. You need to answer what you will do to the observations based on the following information:\\
(1) You are e***1. e***1 is a highly active and influential activist on social media. e***1 enjoys collecting and sharing ideas and opinions, often being the first to do so. e***1 tends to validate, question, challenge, or dismiss the ideas of others and help clarify and steer conversations. e***1's bio includes hashtags and affiliations related to progressive causes such as supporting Joe Biden and Kamala Harris, Black Lives Matter, LGBTQ+ rights, and resistance against oppressive systems. e***1 has retweeted posts about meeting Kamala Harris, hiring Mary to score a film, criticizing Senator Ron Johnson, recommending Neal Katyal for a case, and expressing excitement for the future of Impact.\\
(2) Current time is 2018-01-07 12:00:00\\
(3) The news you got is "At the Golden Globes Awards ceremony in Los Angeles, most guests showed up dressed in black out of solidarity with the MeToo and Time's Up movement and the victims of sexual violence."\\
(4) Your personal experience is e***1 leans towards supporting candidates who prioritize human rights, oppose potential national abortion bans, and criticize some government actions.\\
(5) Your recent memory is [g***n]: g***n replies to [G***s]: I applaud the guests at the Golden Globes for using their platform to support the MeToo and Time's Up movement. It's important to continue raising awareness about sexual violence. \#GoldenGlobes \#MeToo \#TimesUp.\\
{[s***e]: s***e replies to [C***N]: It's disappointing to see President Trump endorsing someone accused of sexual misconduct. We need leaders who take these allegations seriously. \#MeToo.}\\
{[j***3]: j***3 replies to [C***N]: It's concerning that President Trump endorsed Roy Moore despite the allegations of sexual misconduct against him. This sends a message that such behavior is acceptable. \#MeToo \#TimesUp.}\\
{[e***1]: e***1 likes a tweet of [w***r]: 'President Trump's endorsement of Roy Moore, accused of sexual misconduct, is deeply troubling. It undermines the fight against sexual violence and the values of the MeToo movement. \#MeToo \#TimesUp'.}\\
{[T***x]: The solidarity shown at the Golden Globes Awards ceremony in support of the MeToo and Time's Up movement is inspiring. Let's keep the conversation going and work towards a more inclusive and equal society. \#MeToo \#TimesUp}\\
(6) The twitter page you can see is tweet id: 356 [T***x]: The solidarity shown at the Golden Globes Awards ceremony in support of the MeToo and Time's Up movement is inspiring. Let's keep the conversation going and work towards a more inclusive and equal society. \#MeToo \#TimesUp --Post Time: 2018-01-07 04:00:00\\
tweet id: 244 [w***r]: President Trump's endorsement of Roy Moore, accused of sexual misconduct, is deeply troubling. It undermines the fight against sexual violence and the values of the MeToo movement. \#MeToo \#TimesUp --Post Time: 2018-01-06 20:00:00\\
tweet id: 132 [T***x]: I applaud the guests at the Golden Globes for standing in solidarity with the MeToo and Time's Up movement. It's important to support the victims of sexual violence and work towards a safer and more equal society. \#GoldenGlobes \#MeToo \#TimesUp --Post Time: 2018-01-06 20:00:00\\
tweet id: 129 [e***1]: It's inspiring to see the guests at the Golden Globes Awards ceremony showing solidarity with the MeToo and Time's Up movement. This is an important step towards ending sexual violence and creating a safer world for everyone. \#MeToo \#TimesUp \#GoldenGlobes --Post Time: 2018-01-06 20:00:00\\
tweet id: 72 [r***7]: President Trump's endorsement of Roy Moore, despite the allegations of sexual misconduct, is deeply troubling. We must hold our leaders accountable for their actions. \#MeToo \#TimesUp --Post Time: 2018-01-06 20:00:00\\
(7) The notifications you can see are \\
\\
In terms of how you actually perform the action, you take action by calling functions. Currently, there are the following functions that can be called.\\
- do\_nothing(): Do nothing. There is nothing that you like to respond to. \\ 
- post(content): Post a tweet. `content` is the sentence that you will post.  \\
- retweet(content, author, original\_tweet\_id, original\_tweet). Retweet or quote an existing tweet in your Twitter page. `content` is the statements that you attach when retweeting. If you want to say nothing, set `content` to None. `author` is the author of the tweet that you want to retweet, it should be the concrete name. `original\_tweet\_id` and `original\_tweet` are the id and content of the retweeted tweet.  \\
- reply(content, author, original\_tweet\_id). Reply to an existing tweet in your Twitter page or reply one of replies in your notifications, but don't reply to yourself and those not in your Twitter page. `content` is what you will reply to the original tweet or other comments. `author` is the author of the original tweet or comment that you want to reply to. `original\_tweet\_id` is the id of the original tweet.  \\
- like(author, original\_tweet\_id). Press like on an existing tweet in your Twitter page. `author` is the author of the original tweet that you like. `original\_tweet\_id` is the id of the original tweet.  \\
\\
Call one function at a time, please give a thought before calling these actions, i.e., use the following format strictly:\\
\\
{[OPTION 1]}\\
Thought: None of the observation attract my attention, I need to:\\
Action: do\_nothing()\\
\\
{[OPTION 2]}\\
Thought: due to `xxx`, I need to:\\
Action: post(content="yyy")\\
\\
{[OPTION 3]}\\
Thought: due to `xxx`, I need to:\\
Action: retweet(content="yyy", author="zzz", original\_tweet\_id="0", original\_tweet="kkk")\\
\\
{[OPTION 4]}\\
Thought: due to `xxx`, I need to:\\
Action: reply(content="yyy", author="zzz", original\_tweet\_id="0")\\
\\
{[OPTION 5]}\\
Thought: due to `xxx`, I need to:\\
Action: like(author="zzz", original\_tweet\_id="1")\\
\\
Now begin your actions. Remember only write one function call after `Action:`.\\
Based on the above history, what will you, e***1, do next?\\
\end{tcolorbox}
\label{tab:prompt_exp}
\end{center}

\begin{center}
\begin{tcolorbox}[colback=red!5!white,colframe=red!55!black,width=0.95\columnwidth,title={Response Example of Core Users in Metoo Movement.}]
Thought: The observation about the solidarity shown at the Golden Globes in support of the MeToo and Time's Up movement aligns with my progressive values and interests.\\
Action: retweet(content=None, author="T***x", original\_tweet\_id="356", original\_tweet="The solidarity shown at the Golden Globes Awards ceremony in support of the MeToo and Time's Up movement is inspiring. Let's keep the conversation going and work towards a more inclusive and equal society. \#MeToo \#TimesUp")\\
\end{tcolorbox}
\label{tab:response_exp}
\end{center}

\subsection{Parameters of ABMs}~\label{app:calib}

As mentioned in Sec.~\ref{sec:4.4}, we find the best parameter combinations by parameter sweeping on the data of E1 or P1 for each movement, where we run the models with each parameter combination 5 times and record the average result. The parameter combination with the lowest $\Delta Bias$ and $\Delta Div.$  is retained. The parameter combinations used in pure ABMs and our hybrid models are illustrated in Table~\ref{tab:param}. The explanations of each parameter is as follows:
\begin{enumerate}[-]
\itemsep-0.2em
\item alpha:  the strength of the social influence of source agents;
\item bc\_bound: the confidence bound of the agents in the Bounded Confidence Models;
\item init\_uct: the uncertainty term in the Relative Agreement Model;
\item acc\_thred: the threshold for the latitude of acceptance of the agents in the Social Judgement Model;
\item rej\_thred: the threshold for the latitude of rejection of the agents in the Social Judgement Model;
\item lambda: hyperparameter specifying the shape of the similarity bias function in the Lorenz Model; 
\item k: hyperparameter specifying the shape of the similarity bias function in the Lorenz Model; 
\item tho: the degree of assimilation, which controls the relative contribution of the assimilation force versus the reinforcement force in the Lorenz Model.

\end{enumerate}

\begin{table}[!t]
\small
\resizebox{0.48\textwidth}{!}{
\begin{tabular}{@{}ll|cccccccc@{}}
\toprule
\multicolumn{1}{c}{\textbf{Dataset}} & \multicolumn{1}{c|}{\textbf{Models}} & \multicolumn{1}{c}{\textbf{alpha}} & \multicolumn{1}{c}{\textbf{bc\_bound}} & \multicolumn{1}{c}{\textbf{init\_uct}} & \textbf{acc\_thred} & \textbf{rej\_thred} & \textbf{lambda} & \textbf{k}     & \textbf{tho } \\ \midrule
\multirow{5}{*}{Metoo}      & BC                          & 0.10                      & 0.30                         &          -                    &     -      &     -      &    -    &    -   &    -  \\
                            & HK                          & 0.25                      & 0.10                         &             -                 &     -      &       -    &    -    &   -    &   -   \\
                            & RA                          & 0.30                      &          -                    & 0.20                         &       -    &        -   &    -    &    -   &    -  \\
                            & SJ                          & 0.15                      &       -                       &           -                   & 0.10      & 0.90      &   -     &    -   &    -  \\
                            & Lorenz                      & 0.10                      &             -                 &      -                        &     -      &       -    & 1.00   & 2.00  & 0.90 \\ \midrule
\multirow{5}{*}{Roe}        & BC                          & 0.15                      & 0.10                         &             -                 &      -     &        -   &     -   &   -    &    -  \\
                            & HK                          & 0.25                      & 0.10                         &          -                    &      -     &      -     &   -     &    -   &    -  \\
                            & RA                          & 0.10                      &          -                    & 0.20                         &      -     &      -     &    -    &    -   &   -   \\
                            & SJ                          & 0.30                      &  -                            &          -                    & 0.10      & 1.90      &  -      &   -    &   -   \\
                            & Lorenz                      & 0.10                      &   -                           &         -                     &    -       &     -      & 2.00   & 10.00 & 0.50 \\ \midrule
\multirow{5}{*}{BLM}        & BC                          & 0.15                      & 0.10                         &               -               &       -    &       -    &    -    &   -    &    -  \\
                            & HK                          & 0.10                      & 0.10                         &           -                   &       -    &     -      &   -     &   -    &    -  \\
                            & RA                          & 0.10                      &        -                      & 0.20                         &       -    &      -     &      -  &   -    &   -   \\
                            & SJ                          & 0.20                      &  -                            &           -                   & 0.50      & 1.50      & -       &  -     &   -   \\
                            & Lorenz                      & 0.10                      &     -                         &          -                    &     -      &      -     & 2.00   & 2.00  & 0.30 \\ \bottomrule
\end{tabular}}
\caption{Calibrated parameters of the ABMs. ``-'' denotes non-applicable parameters.}
\label{tab:param}
\end{table}

\input{tabs/ablation}

%% file: tabs/ablation.tex
\begin{table}[!t]
\resizebox{0.48\textwidth}{!}{
\begin{tabular}{@{}l|cccccccc@{}}
\toprule
\multirow{2}{*}{\textbf{Datasets}} & \multicolumn{3}{c}{\textbf{Stance}} & \multicolumn{3}{c}{\textbf{Content}} & \multicolumn{2}{c}{\textbf{Behavior}} \\
                  &    Acc.  & F1             & MAE             & Acc.       & F1         & Sim.       & Acc.              & F1                \\ \midrule
Metoo        &     \textbf{0.9679} & \textbf{0.3400 }                & 0.2311          & \textbf{0.7010}     & \textbf{0.1988}     & \textbf{0.8064}     & \textbf{0.7313}            & \textbf{0.5212}            \\
w/o soc.           & 0.9630  &0.2720                & 0.2344          & 0.6671     & 0.1915     & 0.8016     & 0.7019            & 0.5136            \\
w/o com.            & 0.9535 & 0.3393               & \textbf{0.2299}          & 0.6671     & 0.1960     & 0.8027     & 0.7010            & 0.5128            \\ \midrule
Roe             &  \textbf{0.9430} & \textbf{0.3361}                  &      0.2058           &      \textbf{0.6423}      &     \textbf{0.1957}       &     \textbf{0.8090 }      &          \textbf{0.6665}         &     \textbf{0.4691}              \\
w/o soc.   & 0.9302 &0.3214   &      \textbf{0.1914 }              &      0.6120           &    0.1895        &    0.8062        &    0.6403        &       0.4645                       \\
w/o com.      &0.9193 & 0.3229                     &       0.2170          &    0.5839        &    0.1826        &    0.8082        &           0.6106        &     0.4494              \\ \midrule
BLM           & \textbf{0.8991} &\textbf{0.3735}                     & \textbf{0.1627}          & \textbf{0.7353 }    & \textbf{0.2218 }    & \textbf{0.8406 }    & \textbf{0.7796}            & \textbf{0.5759}            \\
w/o soc.      & 0.8679 &0.3692                     & 0.1800          & 0.6220     & 0.1814     & 0.7703     & 0.7570            & 0.5217            \\
w/o com.    &0.8805 & 0.3839                       & 0.1856          & 0.7281     & 0.2075     & 0.8235     & 0.7734            & 0.5663            \\ \bottomrule
\end{tabular}}
\caption{Results of ablation study.}
\label{tab:ablation}
\end{table}

%% file: apps/res.tex
\input{tabs/temp}
\section{Simulation Results}
\subsection{Ablation Study}~\label{app:ablation}
To validate the effectiveness of the proposed components in the LLM agent profile module, we conduct an ablation study on the micro-level alignment of core users, where the social traits (\emph{soc.}) and communication roles (\emph{com.}) are excluded. Table~\ref{tab:ablation} shows that all these elements contribute to the alignment of the users.

\subsection{The Influence of Temperature}
We test different temperatures for the generation of core user agents, to help trade off the certainty and diversity of the generation results. Table~\ref{tab:temp} shows the results of the macro system evaluation when the temperature of LLMs is set to 0 (in our main experiments) and 1. The results show that the temperature parameter influencing the generation content of core user agents can make a huge difference to the systematic results. Among all the candidates,  hybrid models with SJ and Lorenz models show the best robustness on the time series metrics.

\subsection{Visualization of Collective Results}
\begin{figure*}[!t]
    \centering
    \subfigure{
        \includegraphics[width=\textwidth]{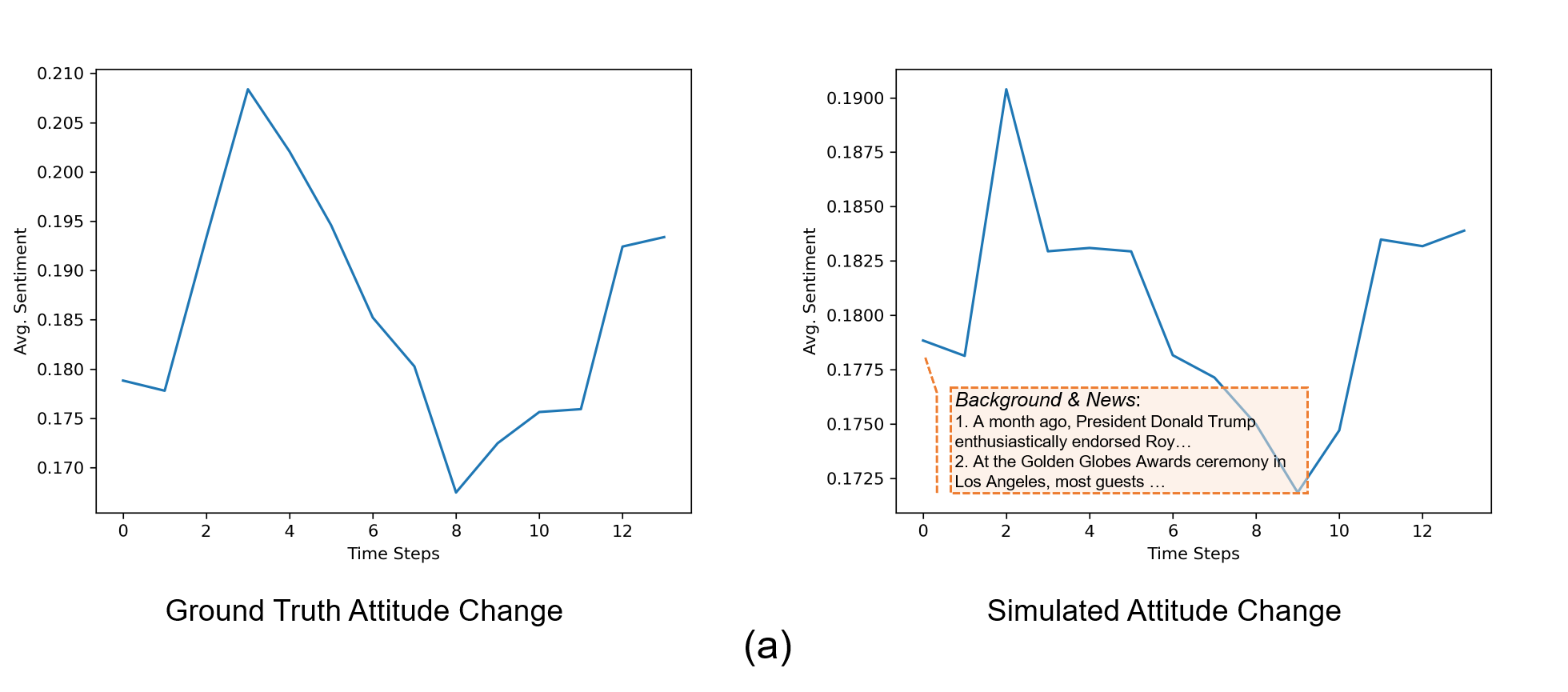}
        \label{fig:metoo_res}
    }
    \vfill
    \subfigure{
        \includegraphics[width=\textwidth]{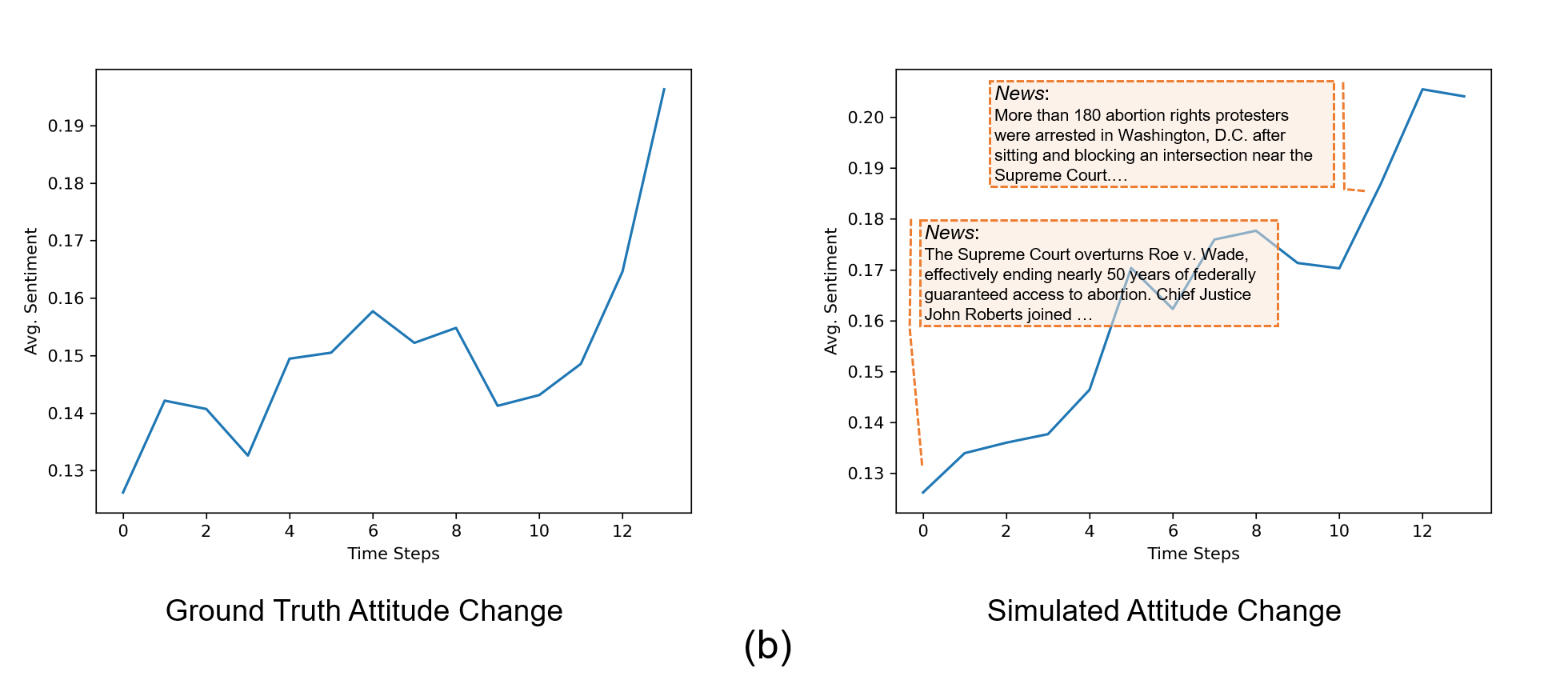}
        \label{fig:roe_res}
    }
    \vfill
    \subfigure{
        \includegraphics[width=\textwidth]{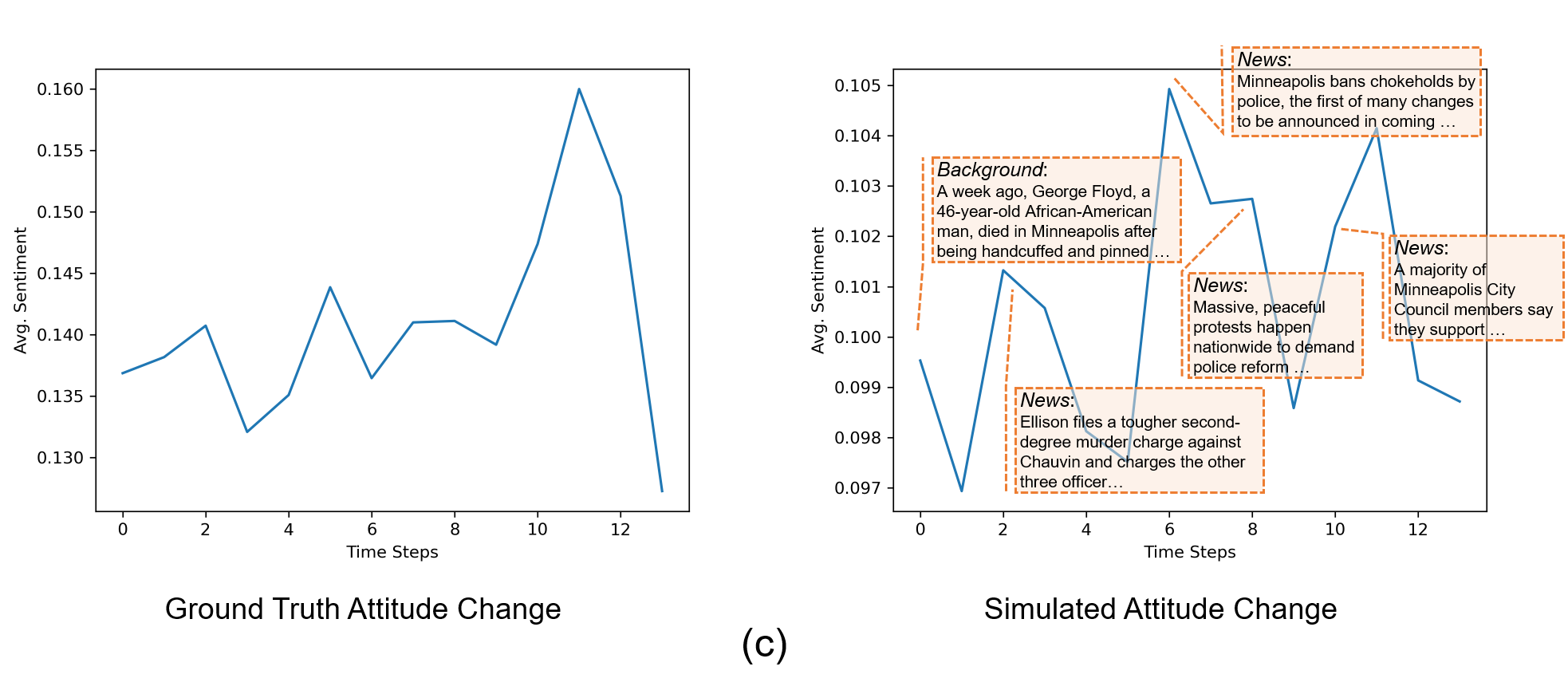}
        \label{fig:blm_res}
    }
    \caption{Examples of collective results. (a) Example of the simulation results of Hybrid w/ BC (temperature=0) on Metoo movement. Since the empirical data contains a discussion about Donald Trump's endorsement of Roy Moore, we also include this background information with the news of the Time's Up movement at the Golden Globes Awards ceremony. (b) Example of the simulation results of Hybrid w/ Lorenz (temperature=1) on Roe movement. (c) Example of the simulation results of Hybrid w/ RA (temperature=0) on BLM movement.
}
    \label{fig:vis_macro}
\end{figure*}

We observe the simulated systematical outcomes at the macro level. Figure~\ref{fig:vis_macro} shows examples of the simulated results and their corresponding true situations.

\subsection{Error Analysis}
In this section, we present the errors and bias in the simulation.
\paragraph{Unintended Behavior}
During the simulation, we observed some unintended behaviors generated by core user agents:
\begin{itemize}
    \item Retweet non-tweet news: sometimes the agents mistake the news as a "tweet" and choose to retweet, although we have emphasized that it is news that can not be retweeted directly. This could be related to the instruction-following ability of LLMs. We believe fine-tuning LLMs with social media user data would mitigate this problem.
    \item Repeat retweeting itself: sometimes the agents repeat retweeting themselves for consecutive steps. This results from the limited context of agents in simulation, where the most severe situation is that there are no people an agent follows in the subgraph, so the signals it receives are limited.
    The root cause of this problem is that the real user context is much more complex than that in simulation, with more information sources from different channels even outside Twitter, so reproducing real behaviors is challenging.
\end{itemize}


%% file: tabs/temp.tex
\begin{table*}[!t]
\resizebox{\textwidth}{!}{
\begin{tabular}{@{}l|c|rrrr|rrrr|rrrr@{}}
\toprule
\multicolumn{1}{c|}{\multirow{2}{*}{\textbf{Method}}} & \multirow{2}{*}{\textbf{temp.}} & \multicolumn{4}{c|}{\textbf{Metoo}}                                                                                                  & \multicolumn{4}{c|}{\textbf{Roe}}                                                                                                    & \multicolumn{4}{c}{\textbf{BLM}}                                                                                                    \\
\multicolumn{1}{c|}{}                                 &                                 & \multicolumn{1}{l}{$\Delta_{Bias}$↓} & \multicolumn{1}{l}{$\Delta_{Div.}$↓} & \multicolumn{1}{l}{DTW↓} & \multicolumn{1}{l|}{Corr.↑} & \multicolumn{1}{l}{$\Delta_{Bias}$↓} & \multicolumn{1}{l}{$\Delta_{Div.}$↓} & \multicolumn{1}{l}{DTW↓} & \multicolumn{1}{l|}{Corr.↑} & \multicolumn{1}{l}{$\Delta_{Bias}$↓} & \multicolumn{1}{l}{$\Delta_{Div.}$↓} & \multicolumn{1}{l}{DTW↓} & \multicolumn{1}{l}{Corr.↑} \\ \midrule
\multirow{2}{*}{Hybrid w/ BC}                         & 0                               & 0.0135                               & 0.0108                               & 1.8440                   & 0.7043                      & 0.0239                               & 0.0121                               & 2.4611                   & 0.3607                      & 0.0300                               & 0.0069                               & 3.9254                   & 0.1248                     \\
                                                      & 1                               & 0.0160                               & 0.0145                               & 3.2916                   & 0.4363                      & 0.0229                               & 0.0030                               & 3.6380                   & 0.2041                      & 0.0290                               & 0.0046                               & 3.5227                   & -0.0838                    \\
\multirow{2}{*}{Hybrid w/ HK}                         & 0                               & 0.0126                               & 0.0037                               & 1.9136                   & 0.6517                      & 0.0319                               & 0.0157                               & 3.6752                   & -0.0807                     & 0.0578                               & 0.0093                               & 3.7288                   & -0.2433                    \\
                                                      & 1                               & 0.0131                               & 0.0091                               & 3.4085                   & 0.3111                      & 0.0334                               & 0.0057                               & 6.1835                   & -0.4999                     & 0.0410                               & 0.0071                               & 4.5660                   & -0.2447                    \\
\multirow{2}{*}{Hybrid w/ RA}                         & 0                               & 0.0117                               & 0.0008                               & 1.7829                   & 0.7238                      & 0.0221                               & 0.0104                               & 2.3326                   & 0.4274                      & 0.0376                               & 0.0070                               & 2.2353                   & 0.6050                     \\
                                                      & 1                               & 0.0116                               & 0.0042                               & 3.1318                   & 0.4829                      & 0.0207                               & 0.0016                               & 2.8546                   & 0.5260                      & 0.0257                               & 0.0035                               & 2.5903                   & 0.6009                     \\
\multirow{2}{*}{Hybrid w/ SJ}                         & 0                               & 0.0098                               & 0.0119                               & 2.2789                   & 0.6327                      & 0.0203                               & 0.0095                               & 1.1896                   & 0.6598                      & 0.0076                               & 0.0018                               & 2.4564                   & 0.5167                     \\
                                                      & 1                               & 0.0107                               & 0.0180                               & 2.9388                   & 0.5447                      & 0.0188                               & 0.0080                               & 2.5737                   & 0.7047                      & 0.0034                               & 0.0037                               & 3.4908                   & 0.5333                     \\
\multirow{2}{*}{Hybrid w/ Lorenz}                     & 0                               & 0.0035                               & 0.0116                               & 2.9857                   & 0.6103                      & 0.0093                               & 0.0147                               & 1.0148                   & 0.7576                      & 0.0023                               & 0.0079                               & 2.5394                   & 0.5055                     \\
                                                      & 1                               & 0.0013                               & 0.0158                               & 2.3384                   & 0.5796                      & 0.0148                               & 0.0052                               & 0.9340                   & 0.8765                      & 0.0154                               & 0.0049                               & 3.2131                   & 0.5038                     \\ \bottomrule
\end{tabular}}
\caption{Results of macro system evaluation with LLM-empowered agents' temperature set to 0 and 1.}
\label{tab:temp}
\end{table*}

%% file: acl2023.bbl
\begin{thebibliography}{47}
\expandafter\ifx\csname natexlab\endcsname\relax\def\natexlab#1{#1}\fi

\bibitem[{Bentwood(2008)}]{bentwooddistributed}
Jonny Bentwood. 2008.
\newblock \href {http://technobabble2dot0.files.wordpress.com/2008/01/edel man-white-paper-distributed-influence-quantifying-the-impact-of-social-media.pdf.} {Distributed influence: Quantifying the impact of social}.

\bibitem[{Br{\"u}nker et~al.(2020)Br{\"u}nker, Wischnewski, Mirbabaie, and Meinert}]{brunker2020role}
Felix Br{\"u}nker, Magdalena Wischnewski, Milad Mirbabaie, and Judith Meinert. 2020.
\newblock The role of social media during social movements--observations from the\# metoo debate on twitter.

\bibitem[{Chang et~al.(2023)Chang, Rao, Zhong, Wojcieszak, and Lerman}]{chang2023roeoverturned}
Rong-Ching Chang, Ashwin Rao, Qiankun Zhong, Magdalena Wojcieszak, and Kristina Lerman. 2023.
\newblock \# roeoverturned: Twitter dataset on the abortion rights controversy.
\newblock In \emph{Proceedings of the International AAAI Conference on Web and Social Media}, volume~17, pages 997--1005.

\bibitem[{Chen et~al.(2023)Chen, Su, Zuo, Yang, Yuan, Qian, Chan, Qin, Lu, Xie et~al.}]{chen2023agentverse}
Weize Chen, Yusheng Su, Jingwei Zuo, Cheng Yang, Chenfei Yuan, Chen Qian, Chi-Min Chan, Yujia Qin, Yaxi Lu, Ruobing Xie, et~al. 2023.
\newblock Agentverse: Facilitating multi-agent collaboration and exploring emergent behaviors in agents.
\newblock \emph{arXiv preprint arXiv:2308.10848}.

\bibitem[{Chuang et~al.(2023)Chuang, Goyal, Harlalka, Suresh, Hawkins, Yang, Shah, Hu, and Rogers}]{chuang2023simulating}
Yun-Shiuan Chuang, Agam Goyal, Nikunj Harlalka, Siddharth Suresh, Robert Hawkins, Sijia Yang, Dhavan Shah, Junjie Hu, and Timothy~T Rogers. 2023.
\newblock Simulating opinion dynamics with networks of llm-based agents.
\newblock \emph{arXiv preprint arXiv:2311.09618}.

\bibitem[{Chuang and Rogers(2023)}]{chuang2023computational}
Yun-Shiuan Chuang and Timothy~T Rogers. 2023.
\newblock Computational agent-based models in opinion dynamics: A survey on social simulations and empirical studies.
\newblock \emph{arXiv preprint arXiv:2306.03446}.

\bibitem[{Cohen et~al.(2009)Cohen, Huang, Chen, Benesty, Benesty, Chen, Huang, and Cohen}]{cohen2009pearson}
Israel Cohen, Yiteng Huang, Jingdong Chen, Jacob Benesty, Jacob Benesty, Jingdong Chen, Yiteng Huang, and Israel Cohen. 2009.
\newblock Pearson correlation coefficient.
\newblock \emph{Noise reduction in speech processing}, pages 1--4.

\bibitem[{Deffuant et~al.(2002)Deffuant, Amblard, Weisbuch, and Faure}]{deffuant2002can}
Guillaume Deffuant, Fr{\'e}d{\'e}ric Amblard, G{\'e}rard Weisbuch, and Thierry Faure. 2002.
\newblock How can extremism prevail? a study based on the relative agreement interaction model.
\newblock \emph{Journal of artificial societies and social simulation}, 5(4).

\bibitem[{Deffuant et~al.(2000)Deffuant, Neau, Amblard, and Weisbuch}]{deffuant2000mixing}
Guillaume Deffuant, David Neau, Frederic Amblard, and G{\'e}rard Weisbuch. 2000.
\newblock Mixing beliefs among interacting agents.
\newblock \emph{Advances in Complex Systems}, 3(01n04):87--98.

\bibitem[{Gao et~al.(2023)Gao, Lan, Lu, Mao, Piao, Wang, Jin, and Li}]{gao2023s}
Chen Gao, Xiaochong Lan, Zhihong Lu, Jinzhu Mao, Jinghua Piao, Huandong Wang, Depeng Jin, and Yong Li. 2023.
\newblock S $^{3}$: Social-network simulation system with large language model-empowered agents.
\newblock \emph{arXiv preprint arXiv:2307.14984}.

\bibitem[{Garimella et~al.(2018)Garimella, De~Francisci~Morales, Gionis, and Mathioudakis}]{garimella2018political}
Kiran Garimella, Gianmarco De~Francisci~Morales, Aristides Gionis, and Michael Mathioudakis. 2018.
\newblock Political discourse on social media: Echo chambers, gatekeepers, and the price of bipartisanship.
\newblock In \emph{Proceedings of the 2018 world wide web conference}, pages 913--922.

\bibitem[{Gestefeld and Lorenz(2023)}]{gestefeld2023calibrating}
Martin Gestefeld and Jan Lorenz. 2023.
\newblock Calibrating an opinion dynamics model to empirical opinion distributions and transitions.
\newblock \emph{Journal of Artificial Societies and Social Simulation}, 26(4).

\bibitem[{Giorgi et~al.(2022)Giorgi, Guntuku, Himelein-Wachowiak, Kwarteng, Hwang, Rahman, and Curtis}]{blmdata}
Salvatore Giorgi, Sharath~Chandra Guntuku, McKenzie Himelein-Wachowiak, Amy Kwarteng, Sy~Hwang, Muhammad Rahman, and Brenda Curtis. 2022.
\newblock Twitter data of the \#blacklivesmatter movement and counter protests: 2013 to 2021.

\bibitem[{Hegselmann et~al.(2002)Hegselmann, Krause et~al.}]{boundedconfidence}
Rainer Hegselmann, Ulrich Krause, et~al. 2002.
\newblock Opinion dynamics and bounded confidence models, analysis, and simulation.
\newblock \emph{Journal of artificial societies and social simulation}, 5(3).

\bibitem[{Hong et~al.(2023)Hong, Zheng, Chen, Cheng, Wang, Zhang, Wang, Yau, Lin, Zhou et~al.}]{hong2023metagpt}
Sirui Hong, Xiawu Zheng, Jonathan Chen, Yuheng Cheng, Jinlin Wang, Ceyao Zhang, Zili Wang, Steven Ka~Shing Yau, Zijuan Lin, Liyang Zhou, et~al. 2023.
\newblock Metagpt: Meta programming for multi-agent collaborative framework.
\newblock \emph{arXiv preprint arXiv:2308.00352}.

\bibitem[{Hua et~al.(2023)Hua, Fan, Li, Mei, Ji, Ge, Hemphill, and Zhang}]{hua2023war}
Wenyue Hua, Lizhou Fan, Lingyao Li, Kai Mei, Jianchao Ji, Yingqiang Ge, Libby Hemphill, and Yongfeng Zhang. 2023.
\newblock War and peace (waragent): Large language model-based multi-agent simulation of world wars.
\newblock \emph{arXiv preprint arXiv:2311.17227}.

\bibitem[{Huang and Carley(2019)}]{huang2019hierarchical}
Binxuan Huang and Kathleen~M Carley. 2019.
\newblock A hierarchical location prediction neural network for twitter user geolocation.
\newblock In \emph{Proceedings of the 2019 Conference on Empirical Methods in Natural Language Processing and the 9th International Joint Conference on Natural Language Processing (EMNLP-IJCNLP)}, pages 4732--4742.

\bibitem[{Jackson et~al.(2017)Jackson, Rand, Lewis, Norton, and Gray}]{jackson2017agent}
Joshua~Conrad Jackson, David Rand, Kevin Lewis, Michael~I Norton, and Kurt Gray. 2017.
\newblock Agent-based modeling: A guide for social psychologists.
\newblock \emph{Social Psychological and Personality Science}, 8(4):387--395.

\bibitem[{Jager and Amblard(2005)}]{socialjudgement}
Wander Jager and Fr{\'e}d{\'e}ric Amblard. 2005.
\newblock Uniformity, bipolarization and pluriformity captured as generic stylized behavior with an agent-based simulation model of attitude change.
\newblock \emph{Computational \& Mathematical Organization Theory}, 10:295--303.

\bibitem[{Jiang et~al.(2023)Jiang, Ren, and Ferrara}]{jiang2023retweet}
Julie Jiang, Xiang Ren, and Emilio Ferrara. 2023.
\newblock Retweet-bert: political leaning detection using language features and information diffusion on social networks.
\newblock In \emph{Proceedings of the International AAAI Conference on Web and Social Media}, volume~17, pages 459--469.

\bibitem[{Liu et~al.(2023)Liu, Yang, Jia, Zhang, Zhou, Dai, Yang, and Vosoughi}]{liu2023training}
Ruibo Liu, Ruixin Yang, Chenyan Jia, Ge~Zhang, Denny Zhou, Andrew~M Dai, Diyi Yang, and Soroush Vosoughi. 2023.
\newblock Training socially aligned language models in simulated human society.
\newblock \emph{arXiv preprint arXiv:2305.16960}.

\bibitem[{Lorenz et~al.(2021)Lorenz, Neumann, and Schr{\"o}der}]{lorenz2021individual}
Jan Lorenz, Martin Neumann, and Tobias Schr{\"o}der. 2021.
\newblock Individual attitude change and societal dynamics: Computational experiments with psychological theories.
\newblock \emph{Psychological Review}, 128(4):623.

\bibitem[{Maiorana et~al.(2020)Maiorana, Morales~Henry, and Weintraub}]{metoodata}
Zachary Maiorana, Pablo Morales~Henry, and Jennifer Weintraub. 2020.
\newblock \href {https://doi.org/10.7910/DVN/2SRSKJ} {{\#metoo Digital Media Collection - Twitter Dataset}}.

\bibitem[{Mou et~al.(2021)Mou, Wei, Chen, Ning, He, Jiang, and Huang}]{mou2021align}
Xinyi Mou, Zhongyu Wei, Lei Chen, Shangyi Ning, Yancheng He, Changjian Jiang, and Xuan-Jing Huang. 2021.
\newblock Align voting behavior with public statements for legislator representation learning.
\newblock In \emph{Proceedings of the 59th Annual Meeting of the Association for Computational Linguistics and the 11th International Joint Conference on Natural Language Processing (Volume 1: Long Papers)}, pages 1236--1246.

\bibitem[{M{\"u}ller(2007)}]{muller2007dynamic}
Meinard M{\"u}ller. 2007.
\newblock Dynamic time warping.
\newblock \emph{Information retrieval for music and motion}, pages 69--84.

\bibitem[{OpenAI(2023)}]{gpt3.5}
OpenAI. 2023.
\newblock \href {https://platform.openai.com/docs/api-reference/chat/create} {Openai gpt3.5-turbo api}.

\bibitem[{Park et~al.(2023)Park, O'Brien, Cai, Morris, Liang, and Bernstein}]{park2023generative}
Joon~Sung Park, Joseph O'Brien, Carrie~Jun Cai, Meredith~Ringel Morris, Percy Liang, and Michael~S Bernstein. 2023.
\newblock Generative agents: Interactive simulacra of human behavior.
\newblock In \emph{Proceedings of the 36th Annual ACM Symposium on User Interface Software and Technology}, pages 1--22.

\bibitem[{Park et~al.(2022)Park, Popowski, Cai, Morris, Liang, and Bernstein}]{park2022social}
Joon~Sung Park, Lindsay Popowski, Carrie Cai, Meredith~Ringel Morris, Percy Liang, and Michael~S Bernstein. 2022.
\newblock Social simulacra: Creating populated prototypes for social computing systems.
\newblock In \emph{Proceedings of the 35th Annual ACM Symposium on User Interface Software and Technology}, pages 1--18.

\bibitem[{Preo{\c{t}}iuc-Pietro and Ungar(2018)}]{preoctiuc2018user}
Daniel Preo{\c{t}}iuc-Pietro and Lyle Ungar. 2018.
\newblock User-level race and ethnicity predictors from twitter text.
\newblock In \emph{Proceedings of the 27th international conference on computational linguistics}, pages 1534--1545.

\bibitem[{Qian et~al.(2023{\natexlab{a}})Qian, Cong, Yang, Chen, Su, Xu, Liu, and Sun}]{qian2023communicative}
Chen Qian, Xin Cong, Cheng Yang, Weize Chen, Yusheng Su, Juyuan Xu, Zhiyuan Liu, and Maosong Sun. 2023{\natexlab{a}}.
\newblock Communicative agents for software development.
\newblock \emph{arXiv preprint arXiv:2307.07924}.

\bibitem[{Qian et~al.(2023{\natexlab{b}})Qian, Dang, Li, Liu, Chen, Yang, Liu, and Sun}]{qian2023experiential}
Chen Qian, Yufan Dang, Jiahao Li, Wei Liu, Weize Chen, Cheng Yang, Zhiyuan Liu, and Maosong Sun. 2023{\natexlab{b}}.
\newblock Experiential co-learning of software-developing agents.
\newblock \emph{arXiv preprint arXiv:2312.17025}.

\bibitem[{Rane and Salem(2012)}]{rane2012social}
Halim Rane and Sumra Salem. 2012.
\newblock Social media, social movements and the diffusion of ideas in the arab uprisings.
\newblock \emph{Journal of international communication}, 18(1):97--111.

\bibitem[{Roy and Goldwasser(2023)}]{roy2023tale}
Shamik Roy and Dan Goldwasser. 2023.
\newblock “a tale of two movements’: Identifying and comparing perspectives in\# blacklivesmatter and\# bluelivesmatter movements-related tweets using weakly supervised graph-based structured prediction.
\newblock In \emph{Findings of the Association for Computational Linguistics: EMNLP 2023}, pages 10437--10467.

\bibitem[{Schelling(2006)}]{schelling2006micromotives}
Thomas~C Schelling. 2006.
\newblock \emph{Micromotives and macrobehavior}.
\newblock WW Norton \& Company.

\bibitem[{Schwartz et~al.(2013)Schwartz, Eichstaedt, Kern, Dziurzynski, Ramones, Agrawal, Shah, Kosinski, Stillwell, Seligman et~al.}]{schwartz2013personality}
H~Andrew Schwartz, Johannes~C Eichstaedt, Margaret~L Kern, Lukasz Dziurzynski, Stephanie~M Ramones, Megha Agrawal, Achal Shah, Michal Kosinski, David Stillwell, Martin~EP Seligman, et~al. 2013.
\newblock Personality, gender, and age in the language of social media: The open-vocabulary approach.
\newblock \emph{PloS one}, 8(9):e73791.

\bibitem[{Sherif and Hovland(1961)}]{sherif1961social}
Muzafer Sherif and Carl~I Hovland. 1961.
\newblock Social judgment: Assimilation and contrast effects in communication and attitude change.

\bibitem[{Tinati et~al.(2012)Tinati, Carr, Hall, and Bentwood}]{tinati2012identifying}
Ramine Tinati, Leslie Carr, Wendy Hall, and Jonny Bentwood. 2012.
\newblock Identifying communicator roles in twitter.
\newblock In \emph{Proceedings of the 21st international conference on World Wide Web}, pages 1161--1168.

\bibitem[{T{\"o}rnberg et~al.(2023)T{\"o}rnberg, Valeeva, Uitermark, and Bail}]{tornberg2023simulating}
Petter T{\"o}rnberg, Diliara Valeeva, Justus Uitermark, and Christopher Bail. 2023.
\newblock Simulating social media using large language models to evaluate alternative news feed algorithms.
\newblock \emph{arXiv preprint arXiv:2310.05984}.

\bibitem[{Wang et~al.(2023{\natexlab{a}})Wang, Xie, Jiang, Mandlekar, Xiao, Zhu, Fan, and Anandkumar}]{wang2023voyager}
Guanzhi Wang, Yuqi Xie, Yunfan Jiang, Ajay Mandlekar, Chaowei Xiao, Yuke Zhu, Linxi Fan, and Anima Anandkumar. 2023{\natexlab{a}}.
\newblock Voyager: An open-ended embodied agent with large language models.
\newblock \emph{arXiv preprint arXiv:2305.16291}.

\bibitem[{Wang et~al.(2023{\natexlab{b}})Wang, Ma, Feng, Zhang, Yang, Zhang, Chen, Tang, Chen, Lin et~al.}]{wang2023survey}
Lei Wang, Chen Ma, Xueyang Feng, Zeyu Zhang, Hao Yang, Jingsen Zhang, Zhiyuan Chen, Jiakai Tang, Xu~Chen, Yankai Lin, et~al. 2023{\natexlab{b}}.
\newblock A survey on large language model based autonomous agents.
\newblock \emph{arXiv preprint arXiv:2308.11432}.

\bibitem[{Wang et~al.(2023{\natexlab{c}})Wang, Zhang, Chen, Lin, Song, Zhao, and Wen}]{wang2023recagent}
Lei Wang, Jingsen Zhang, Xu~Chen, Yankai Lin, Ruihua Song, Wayne~Xin Zhao, and Ji-Rong Wen. 2023{\natexlab{c}}.
\newblock Recagent: A novel simulation paradigm for recommender systems.
\newblock \emph{arXiv preprint arXiv:2306.02552}.

\bibitem[{Wu et~al.(2024)Wu, Mou, Xue, Ying, Wang, Zhang, Huang, and Wei}]{wu2024pasum}
Kun Wu, Xinyi Mou, Lanqing Xue, Zhenzhe Ying, Weiqiang Wang, Qi~Zhang, Xuan-Jing Huang, and Zhongyu Wei. 2024.
\newblock Pasum: A pre-training architecture for social media user modeling based on text graph.
\newblock In \emph{Proceedings of the 2024 Joint International Conference on Computational Linguistics, Language Resources and Evaluation (LREC-COLING 2024)}, pages 12644--12656.

\bibitem[{Xi et~al.(2023)Xi, Chen, Guo, He, Ding, Hong, Zhang, Wang, Jin, Zhou et~al.}]{xi2023rise}
Zhiheng Xi, Wenxiang Chen, Xin Guo, Wei He, Yiwen Ding, Boyang Hong, Ming Zhang, Junzhe Wang, Senjie Jin, Enyu Zhou, et~al. 2023.
\newblock The rise and potential of large language model based agents: A survey.
\newblock \emph{arXiv preprint arXiv:2309.07864}.

\bibitem[{Xiao et~al.(2020)Xiao, Song, Xu, Ren, and Sun}]{xiao2020timme}
Zhiping Xiao, Weiping Song, Haoyan Xu, Zhicheng Ren, and Yizhou Sun. 2020.
\newblock Timme: Twitter ideology-detection via multi-task multi-relational embedding.
\newblock In \emph{Proceedings of the 26th ACM SIGKDD international conference on knowledge discovery \& data mining}, pages 2258--2268.

\bibitem[{Xu et~al.(2023)Xu, Wang, Li, Luo, Wang, Liu, and Liu}]{xu2023exploring}
Yuzhuang Xu, Shuo Wang, Peng Li, Fuwen Luo, Xiaolong Wang, Weidong Liu, and Yang Liu. 2023.
\newblock Exploring large language models for communication games: An empirical study on werewolf.
\newblock \emph{arXiv preprint arXiv:2309.04658}.

\bibitem[{Zhang et~al.(2023)Zhang, Sheng, Chen, Li, Deng, Wang, and Chua}]{zhang2023generative}
An~Zhang, Leheng Sheng, Yuxin Chen, Hao Li, Yang Deng, Xiang Wang, and Tat-Seng Chua. 2023.
\newblock On generative agents in recommendation.
\newblock \emph{arXiv preprint arXiv:2310.10108}.

\bibitem[{Zhou et~al.(2023)Zhou, Zhu, Mathur, Zhang, Yu, Qi, Morency, Bisk, Fried, Neubig et~al.}]{zhou2023sotopia}
Xuhui Zhou, Hao Zhu, Leena Mathur, Ruohong Zhang, Haofei Yu, Zhengyang Qi, Louis-Philippe Morency, Yonatan Bisk, Daniel Fried, Graham Neubig, et~al. 2023.
\newblock Sotopia: Interactive evaluation for social intelligence in language agents.
\newblock \emph{arXiv preprint arXiv:2310.11667}.

\end{thebibliography}
